\documentclass[twocolumn,showpacs,preprintnumbers,prl,superscriptaddress]{revtex4-2}

\usepackage{hyperref}
\usepackage{amsmath}
\usepackage{color}
\usepackage{txfonts}
\usepackage{graphicx}
\usepackage{wrapfig}
\usepackage{comment}
\usepackage{bm}
\usepackage{here}
\usepackage{widetable}
\usepackage[switch,columnwise]{lineno}
\usepackage{dcolumn}
\usepackage{natbib}

\usepackage{natbib}
\usepackage{lineno}

\begin{document}
\title{Pressure-induced nearly perfect rectangular lattice and superconductivity in an organic molecular crystal
(DMET-TTF)$_2$AuBr$_2$}
\author{Taiga Kato}
\affiliation{Department of Physics, Hokkaido University, Sapporo 060-0810, Japan}
\author{Hanming Ma}
\affiliation{Institute for Solid State Physics (ISSP), University of Tokyo, Kashiwa, Chiba 277-8581, Japan}
\author{Kazuyoshi Yoshimi}
\affiliation{Institute for Solid State Physics (ISSP), University of Tokyo, Kashiwa, Chiba 277-8581, Japan}
\author{Takahiro Misawa}
\affiliation{Institute for Solid State Physics (ISSP), University of Tokyo, Kashiwa, Chiba 277-8581, Japan}
\author{Shigen Kumagai}
\author{Youhei Iida}
\author{Yoshiaki Sasaki}
\author{Masashi Sawada}
\affiliation{Department of Physics, Hokkaido University, Sapporo 060-0810, Japan}
\author{Jun Gouchi}
\affiliation{Institute for Solid State Physics (ISSP), University of Tokyo, Kashiwa, Chiba 277-8581, Japan}
\author{Takuya Kobayashi}
\author{Hiromi Taniguchi}
\affiliation{Graduate School of Science and Engineering, Saitama University, Shimo-Ohkubo 255, Saitama 338-8570, Japan}
\author{Yoshiya Uwatoko}
\affiliation{Institute for Solid State Physics (ISSP), University of Tokyo, Kashiwa, Chiba 277-8581, Japan}
\author{Hiroyasu Sato}
\affiliation{Rigaku Corporation, Akishima, 196-8666, Japan}
\author{Noriaki Matsunaga}
\email{E-mail:mat@phys.sci.hokudai.ac.jp}
\affiliation{Department of Physics, Hokkaido University, Sapporo 060-0810, Japan}
\author{Atsushi Kawamoto}
\author{Kazushige Nomura}
\affiliation{Department of Physics, Hokkaido University, Sapporo 060-0810, Japan}

\date{\today}

%TC:ignore
%ABSTRACT%
\begin{abstract}
{
External pressure and associated changes in lattice structures are key  to realizing exotic quantum phases such as
 high-$T_{\rm c}$ superconductivity. While applying external pressure is a 
standard method to induce 
novel lattice structures, its impact on organic molecular crystals has been less explored. Here we report a unique structural phase transition in (DMET-TTF)$_2$AuBr$_2$ under pressure. 
By combining advanced high-pressure techniques and $ab$ $initio$ calculations, we elucidate that (DMET-TTF)$_2$AuBr$_2$ undergoes a transition from a quasi-one-dimensional lattice to a nearly perfect rectangular lattice at 0.9 GPa. This transition leads to the realization of an antiferromagnetic Mott insulator with $T_{\rm N}=66$ K, the highest $T_{\rm N}$ in low-dimensional molecular crystal solids to date. 
Upon increasing the pressure, the antiferromagnetic ordering is suppressed, and a superconducting phase with $T_{\rm c}=4.8$ K emerges around 6 GPa. Our study reveals the significant impact of external pressure on lattice structures of organic molecular crystals and highlights the intricate relationship between 
geometrical frustration and superconductivity. 
Our findings also pave the way for realizing 
functional organic molecular crystals 
through changes in lattice structures by pressure.
}
\end{abstract}
\maketitle

{\it Introduction---.}
External pressure has been used to realize exotic quantum phases in solids such
as the high-$T$ superconductivity since it can cause 
significant changes in electronic structures via changes in lattice 
%constants and/or 
structures.
%of solids.
In the study of the high-$T_{\rm c}$ superconductivity,
it has been shown that external pressure often raises $T_{\rm c}$~\cite{Gao_PRB1994,Monteverde_EPL2005,Takeshita_JPSJ2013,Takahashi_Nature2008,Stewart_RMP2011,Wang_NCom2022,Sun_Nature2023,Wang_PRX2024}.
For example, the highest $T_{\rm c}$ of the cuprates ($T_{\rm c}\sim 166$ K)~\cite{Gao_PRB1994,Monteverde_EPL2005,Takeshita_JPSJ2013} has been 
found under pressure.
The discovery of high-$T_{\rm c}$ superconductivity above $200$ K in hydrogen sulfides~\cite{SCH2S,drozdov2019superconductivity} 
under high pressure is groundbreaking because these materials exhibit critical temperatures close to room temperature.

Organic molecular crystals are known for their high sensitivity to external pressures due to their softness.
In organic molecular crystals, it has been shown that the pressure induces various exotic quantum phases, such as a spin-Peierls phase, a charge-ordered phase, and a superconducting phase~\cite{TMTCF,organic_superconductors}.
However, there are only a few reports of pressure-induced {\it structural phase 
transitions} in organic molecular crystals~\cite{TM_st_transition}.
Moreover, it has been a challenging problem
to identify structural phase transitions
in organic molecular solids under high pressure using 
a diamond anvil cell (DAC)~\cite{organic_xray}, 
due to the complexity of their crystal structures and the small sample sizes.

In this paper, we report 
a pressure-induced structural phase transition in the organic conductor (DMET--TTF)$_2$AuBr$_2$~\cite{AuBr2_AP_C-SDW},
where DMET--TTF is an abbreviation for 
dimethyl(ethylenedithio)tetrathiafulvalene.
This structural phase transition is caused by the donor arrangement. At ambient pressure, this compound is a quasi-one-dimensional (Q1D) electronic system
\cite{AuBr2_AP_C-SDW}.
Upon applying pressure, a pressure-induced structural phase transition occurs around 0.9 GPa.
Just after this structural phase transition, we find that an antiferromagnetic (AF) transition occurs 
around $T_{\rm N}=66$ K,
which is 
the highest $T_{\rm N}$ 
among all low-dimensional organic conductors. 
$Ab$ $initio$ calculations indicate that the 
low-energy effective model after the structural phase transition
is the extended Hubbard model on a nearly perfect rectangular lattice, where geometrical frustration is almost absent.
This result is in sharp contrast to other low-dimensional organic molecular solids, 
which often form frustrated lattices such as an anisotropic triangular lattice~\cite{Kanoda_ARCMP2011,Powell_review2006}.

\begin{figure*}[htb]
\centering 
\includegraphics[width=180mm]{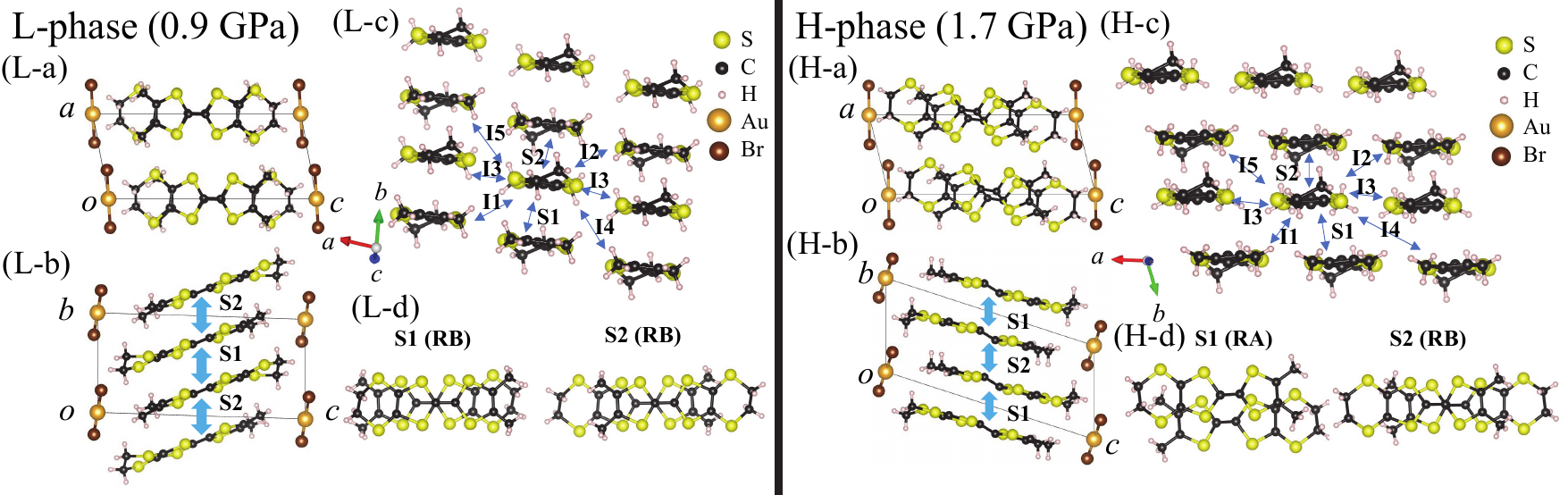}
\caption{
Left panel: crystal structure of the low-pressure phase (L-phase) of (DMET--TTF)$_2$AuBr$_2$ at a pressure of 0.9 GPa. 
(L-a) As viewed along the $b$-axis, 
(L-b) viewed along the $a$-axis, 
and 
(L-c) 
viewed along the long axis of the DMET-TTF molecule.
(L-d) Overlaps of the DMET--TTF molecules viewed from the direction normal to the molecular plane. 
Right panel: crystal structure of the high-pressure phase (H-phase) of (DMET--TTF)$_2$AuBr$_2$ at a pressure of 1.7 GPa. 
(H-a) As viewed along the $b$-axis, 
(H-b) viewed along the $a$-axis, 
(H-c) viewed along the long axis of the DMET-TTF molecule,
and 
(H-d) overlaps of the DMET--TTF molecules viewed from the direction normal to the molecular plane.}

\label{St}
\end{figure*}

When the pressure is increased further to around $6$ GPa, the compound undergoes a phase transition from an AF phase to a superconducting (SC) phase with an optimal SC transition temperature
$T_{\rm c}^{\rm opt}=4.8$ K.
The appearance of an SC phase adjacent to an AF phase is a common behavior in high-$T_{\rm c}$ compounds such as the cuprates, the iron-based superconductors, and the organic salts~$\mathrm{\kappa}$-BEDT-TTF \cite{kET_AF,k-Br}.
A SC transition temperature 
adjacent to an antiferromagnetic phase 
is expected to scale with the energy scale of the AF interactions; i.e., 
with the AF transition temperature~\cite{YBCO}. 
However, the optimal SC transition temperature in (DMET--TTF)$_2$AuBr$_2$ is 
\t,{significantly} lower than the highest SC transition temperature found 
in $\mathrm{\beta}'$-(BEDT-TTF)$_2$ICl$_2$~\cite{ICl2_structure,ICl2_Taniguchi} ($T_{\rm c}^{\rm opt}\sim$ $14.2$ K at $8.2$ GPa),
whereas the maximum AF transition temperature is 
$50$ K at 2.5GPa~\cite{ICl2_TN,Icl2_NMR}.
This result suggests that geometrical frustration is 
an important factor in stabilizing the superconductivity since 
it suppresses the competing AF phases.

{\it Pressure-induced structural phase transition
---.}
The crystal structure of (DMET-TTF)$_2$AuBr$_2$ at 0.9 GPa is shown in Figs. \ref{St}(L-a), (L-b), and (L-c).
This structure is the same as that observed at ambient pressure \cite{AuBr2_AP_C-SDW}; 
the space group of this structure is $P\bar{1}$, and the DMET--TTF molecules are stacked alternately along the $b$-axis. 
The crystal structure of (DMET-TTF)$_2$AuBr$_2$ at 1.7 GPa is shown in Figs. \ref{St}(H-a), (H-b), and (H-c). 
The space group of this structure also is  $P\bar{1}$, but it is significantly different from the structure observed at 0.9 GPa. 
Details of
the lattice parameters 
are shown in Table S1 in the Supplemental Material~\cite{supplementary}.
Hereafter, we refer to the crystal structure at low pressure as the \lq\lq L-phase\rq\rq\ and to that observed at high pressure as the \lq\lq H-phase\rq\rq.

Here, we describe how to characterize the differences between the lattice structures of the L-phase and the H-phase.
According to the definition 
by Mori \cite{RARB}, the transfer integrals S1 and S2 in the L-phase and S2 in the H-phase are classified as ring-over-bond (RB) overlap modes, and S1 in the H-phase is classified as a ring-over-atom (RA) overlap mode. 
As shown in Figs. \ref{St}(L-d) and (H-d), RA and RB overlap modes are stacked alternately along the $b$-axis in the H-phase, whereas all overlap modes are RB in the L-phase. 
This alternately stacking, previously unobserved in other DMET compounds, 
represents a unique H-phase structure not stabilized by chemical substitution at ambient pressure.

{\it $^{13}$C-NMR measurements under pressure
---.}To investigate the electronic 
structure of the H-phase, 
we synthesized single-site $^{13}$C-substituted DMET--TTF 
molecules via the cross-coupling method 
~\cite{coupling_1,coupling_2}~.
The single NMR peak splits below 61 K at a pressure of 1.0 GPa and below 66 K at a pressure of 1.5 GPa. 
The spin-lattice relaxation rate, $T_1^{-1}$,
diverges due to critical slowing down at the same temperature. 
This behavior provides strong evidence of an AF transition. 
These AF transition temperatures are rather high AF transition temperatures for a low-dimensional organic conductor when compared with 27 K 
for $\mathrm{\kappa}$-(BEDT--TTF)$_2$Cu[N(CN)$_2$]Cl \cite{kET_AF} 
and 22 K for $\mathrm{\beta}'$-(BEDT--TTF)$_2$ICl$_2$ \cite{ICl2_TN}.

\begin{figure*}[t]
\centering 
\includegraphics[width=160mm]{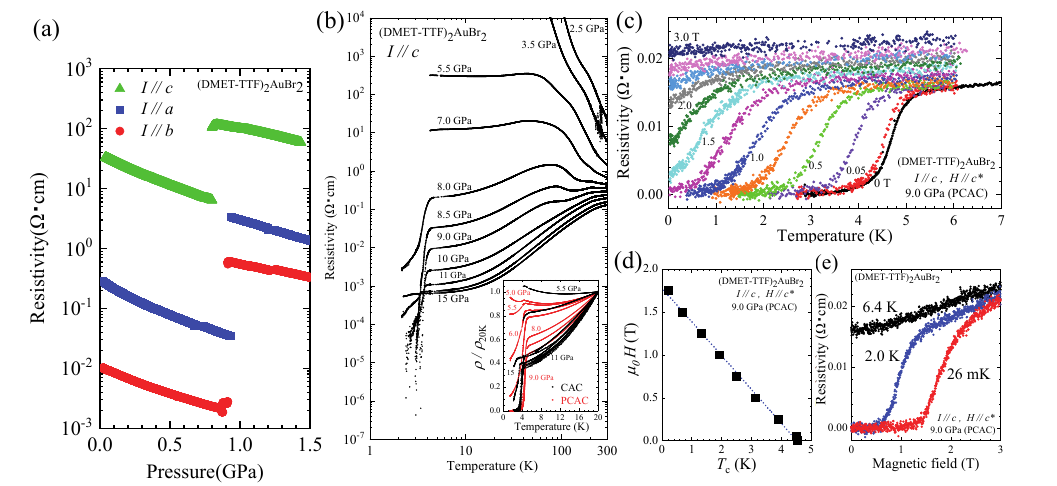}
\caption{
(a) Pressure dependence of the resistivity along the $a$-, $b$-, and $c$-axes at room temperature. 
Around 0.9 GPa, the resistivities along the $a$-, $b$-, and $c$-axes jump discontinuously by factors of 320, 96, and 18, respectively. 
(b) Temperature dependence of the $c$-axis resistivity for pressures up to 15 GPa. 
The black symbols represent data obtained using a CAC. 
The inset shows the resistivities normalized at 20K, $\rho/\rho_{\rm 20 K}$. 
c) Temperature dependence of the $c$-axis resistivity at 9.0 GPa (PCAC); from bottom to top, the measurements were obtained at the magnetic-field strengths 0, 0.05, 0.25, 0.5, 0.75, 1.0, 1.25, 1.5, 1.75, 2.0, 2.25, 2.5, and 3.0 T. 
(d) Magnetic field dependence of $T_{\rm c}$,
where $T_{\rm c}$ is defined as the temperature at which the $c$-axis resistivity at 9.0 GPa (PCAC) becomes half of the residual resistivity.
The blue dotted line shows a linear fit.
(e) Magnetic-field dependence of the $c$-axis resistivity
(PCAC).
}
\label{Anisotropy}
\end{figure*}

{\it Resistivity measurements under pressure and the superconducting transition---.}
Figure \ref{Anisotropy}(a) shows the resistivities along the $a$-, $b$-, 
and $c$-axes at room temperature plotted as a function of the pressure.
The resistivities decrease continuously with increasing pressure
up to an applied pressure of 0.8 GPa.
A discontinuous increase occurs around 0.9 GPa, which 
is consistent with
the structural phase transition.
Above 1.0 GPa, the resistivity again decreases continuously.
At a pressure of around 0.9 GPa, the in-plane anisotropy of the resistivity, $\rho_a$/$\rho_b$, decreases discontinuously from $\rho_a$/$\rho_b = 18$ to $5.6$ (Fig.~5 (a) in the Supplemental Material).
Thus, a large decrease in the anisotropy of the resistivity occurs at the aforementioned resistivity discontinuity. 
This implies that the electronic state undergoes a transition from a Q1D electronic state to a quasi-two-dimensional (Q2D) state at the resistance discontinuity. 
These behaviors are consistent with the structural phase transition determined by the
x-ray diffraction measurements.

To investigate the existence of the SC phases adjacent to the AF phase,
we perform resistivity measurements up to around 15 GPa.
We applied the pressure
using a clamp-type palm cubic anvil cell (PCAC) \cite{PCAC} for pressures up to 9.0 GPa and
using a constant-loading-type cubic anvil cell (CAC) \cite{CAC} for pressures up to 15 GPa. 
Figure \ref{Anisotropy}(b) shows the temperature dependence of the $c$-axis resistivity for pressures up to 15 GPa. 
In the H-phase, 
the resistivity shows insulator-like behavior across the whole temperature range investigated here for pressures up to 3.5 GPa. 
This result suggests that the electronic state in the H-phase is that of a dimer Mott insulator. 
Above 5.5 GPa, however, the resistivity has a broad maximum above $\sim$50 K. 
This broad maximum is not present for pressures above 9.0 GPa, 
and the insulating behavior is suppressed at these pressures; in these conditions (DMET--TTF)$_2$AuBr$_2$ shows metallic behavior even at high temperatures.
Above 5.0 GPa, there is a rapid decrease in the resistivity at $\sim$4 K for pressures up to 15 GPa. 
As shown in the inset of Fig. \ref{Anisotropy}(b), up to 6.0 GPa, the resistivity increases slightly before a rapid decrease occurs with decreasing temperature.
Between 8.5 and 11 GPa, we observed zero resistivity. 
For example, at 9.0 GPa (using the PCAC), the zero-resistivity region extends up to a maximum of 3.42 K. 
Here, we define $T_{\rm c}$ (onset) as the temperature at which a linear fit to the resistivity as a function of temperature in the region above the transition temperature intersects with a similar fit in the region below the transition temperature; for a pressure of 9.0 GPa (using the PCAC), this yields $T_{\rm c}$ (onset) = 4.8 K.

When we applied a magnetic field normal to the conducting plane at 9.0 GPa, the value of $T_{\rm c}$ decreased monotonically, as shown in Fig. \ref{Anisotropy}(c). 
Using the Ginzburg--Landau (GL) theory \cite{organic_superconductors}, we find the GL coherence length, $\xi_{\parallel}$, to be 136 $\AA$. We also find 
$[d\mu_0H_{\rm c2}(T)/dT]_{T_{\rm c}} = -0.395$ T/K and $\mu_0H_{\rm c2}(0)=1.78$ T, 
as determined from the linear fit to $H_{\rm c2}(T)$ shown in Fig. \ref{Anisotropy}(d). 
The decrease in resistivity disappears for applied magnetic fields above 2.5 T, as shown in Fig. \ref{Anisotropy}(e). 
We, therefore, conclude that this phase transition is a SC transition. 
To date, (DMET--TTF)$_2$AuBr$_2$ has the highest value of $T_{\rm c}$ in the DMET family of materials \cite{organic_superconductors}. 

\begin{figure}[b]
\centering
\includegraphics[width=70mm]{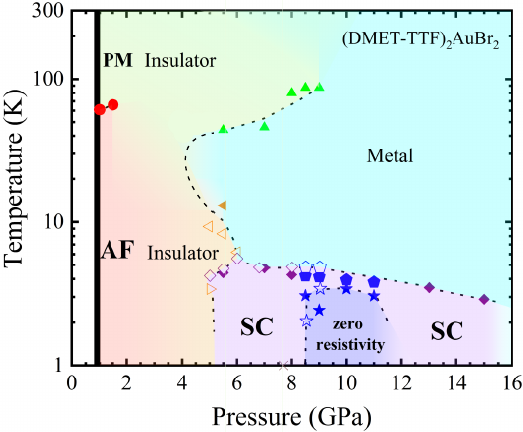}
\caption{
The $P-T$ phase diagram of (DMET--TTF)$_2$AuBr$_2$ in the H-phase. 
The red circles represent the values of $T_{\rm N}$ obtained from the $^{13}$C-NMR measurements. 
Above 2 GPa, the closed symbols represent data using the CAC, and the open symbols represent data using the PCAC. 
The triangular symbols show metal-insulator transition points, defined as the points at which a maximum or minimum of the resistivity 
is observed in the temperature dependence of the $c$-axis resistivity. 
The purple diamonds represent the temperatures at which the resistivity begins to decrease rapidly, the blue pentagons represent $T_{\rm c}$ (onset), and the blue stars show the zero-resistivity point.
}
\label{Diagram}
\end{figure}

\begin{figure*}[t] 
\begin{center} 
\includegraphics[width=160mm]{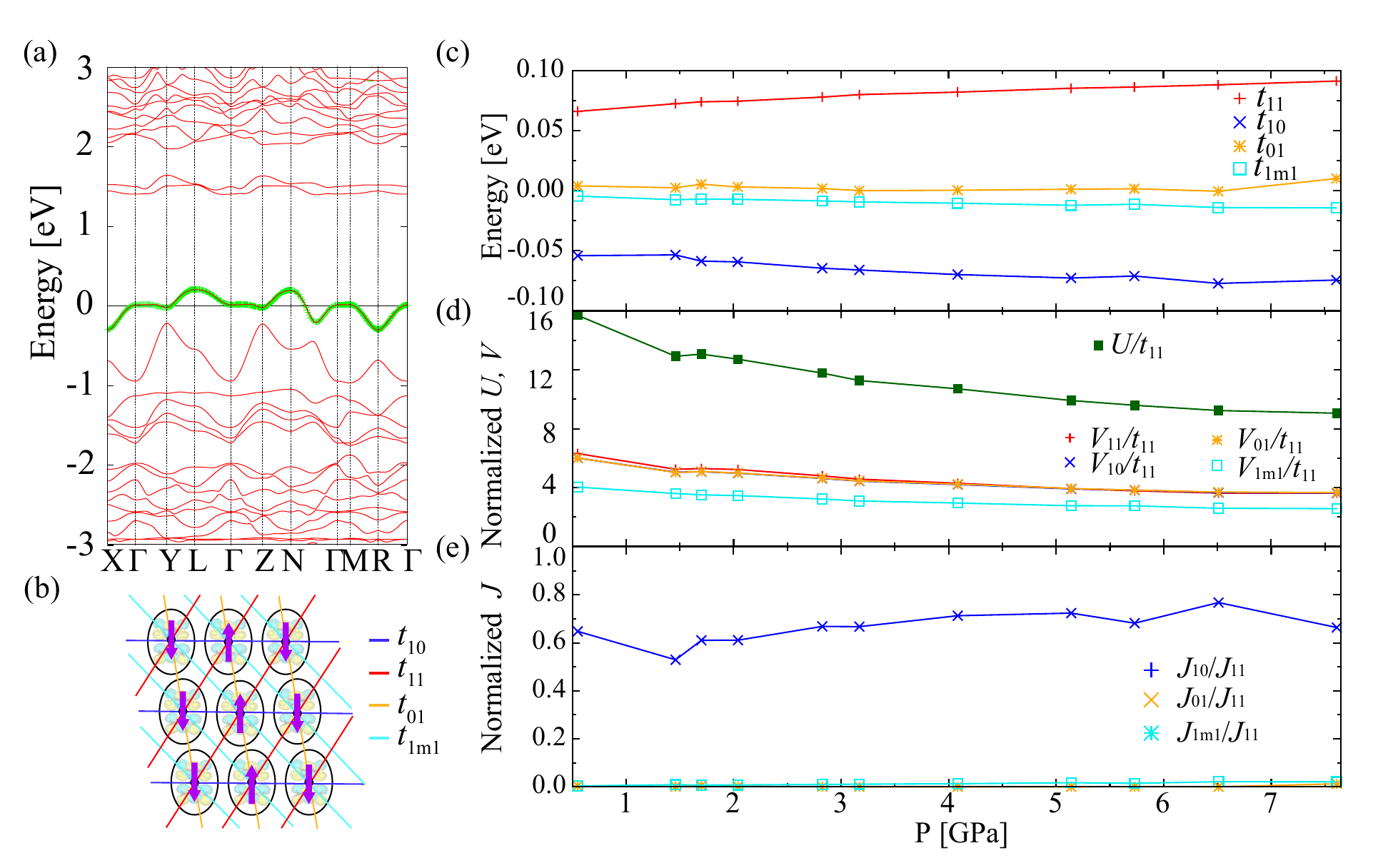}
\caption{
(a) Band structure of (DMET--TTF)$_2$AuBr$_2$ for the strucute at 
$1.7$ GPa. The thin solid (red) lines represent band structures obtained by the density-functional theory (DFT). The bold (green) lines are band structures obtained by 
the maximally localized Wannier functions (MLWFs). We set the Fermi energy to zero (the solid line). Here, we define following indices: X=$(0, -\pi, 0)$, $\Gamma=(0,0, 0)$, Y=$(\pi, 0, 0)$, L=$(\pi, -\pi, 0)$, Z=$(-\pi, 0, \pi)$, N=$(-\pi, -\pi, \pi)$, M=$(0, 0, \pi)$, and R=$(0, -\pi, \pi)$. 
(b) A schematic model of (DMET--TTF)$_2$AuBr$_2$ in the dimer picture. 
The expected antiferromagnetic order is represented by arrows.
(c) The pressure dependence of the transfer integrals between dimers: $t_{11}$, $t_{10}$, $t_{01}$ and $t_{1{\rm m}1}$. (d) Pressure dependence of the normalized on-site screened-Coulomb interactions on a dimer, $U/t_{11}$, and of neighboring screened-Coulomb interactions between dimers, $V_{11}/t_{11}$, $V_{10}/t_{11}$, $V_{01}/t_{11}$, and $V_{1{\rm m}1}/t_{11}$. (e) Pressure dependence of the normalized magnetic exchange--correlation interactions $J_{10}/J_{11}$, $J_{01}/J_{11}$ and $J_{1{\rm m}1}/J_{11}$, where $J_{ij}\equiv 4t_{ij}^2/(U-V_{ij})$.}
\label{fig-params}
\end{center}
\end{figure*}

Figure \ref{Diagram} shows the phase diagram of (DMET--TTF)$_2$AuBr$_2$ in the H-phase, as determined from the resistivity [Fig. \ref{Anisotropy}] and $^{13}$C-NMR measurements (Fig.~6 in the Supplemental Material). 
(DMET-TTF)$_2$AuBr$_2$ exhibits a SC phase over a wide range of pressures. 
The SC phase exists near the AF insulating phase, as is seen in other organic superconductors \cite{k-Br,ICl2_Taniguchi}.
The phase structure of the metal-insulator (M-I) boundary and the Fermi-liquid-like behavior just above $T_{\rm c}$ (Fig.~5(c) in the Supplemental Material) are also similar to observations of the $\mathrm{\kappa}$-BEDT-TTF salts \cite{k-Br}.
The M--I transition in the phase diagram of the $\mathrm{\kappa}$-BEDT-TTF salts has been discussed as a quantum-critical phenomenon in references \cite{Kanoda_ARCMP2011,Kagawa_Nature2005,QCP1,QCP2}
However, geometrical frustration is almost absent in (DMET-TTF)$_2$AuBr$_2$, allowing us to investigate how it affects quantum critical phenomena.

{\it $Ab$ $initio$ effective Hamiltonians---.}
We have performed $ab$ $initio$ calculations to clarify the electronic structure of the H-phase from a 
microscopic viewpoint. 
We first construct the maximally localized Wannier function (MLWF) for describing the band near the Fermi level.
We confirmed that the MLWFs well reproduce the bands near the Fermi level obtained from the density-functional theory (DFT) calculations as shown in {Fig. \ref{fig-params} (a).
Next, we obtained the low-energy effective Hamiltonian in the H-phase, which 
is given by
\begin{equation}
H=\sum_{i,j,\sigma}t_{ij}c_{i\sigma}^{\dagger}c_{j\sigma}
+U\sum_{i}n_{i\uparrow}n_{i\downarrow}
+\sum_{i,j,\sigma}V_{ij}N_{i}N_{j},
\end{equation}
where $i$ ($\sigma$) denotes the site index (spin index), and
the number operators are defined as 
 $n_{i\sigma}=c_{i\sigma}^{\dagger}c_{i\sigma}$ and
 $N_{i}=n_{i\uparrow}+n_{i\downarrow}$.
We evaluated the hopping integrals $t_{ij}$, the on-site Coulomb
interaction $U$, and the off-site Coulomb interactions $V_{ij}$
using the MLWFs and the constrained random phase approximation (cRPA)~\cite{PhysRevB.70.195104,Imada_JPSJ2010,RESPACK}.

We can map the original two-dimensional conducting layer
into the equivalent two-dimensional square lattice, as shown in Fig. \ref{fig-params} (b).
Transfer integrals and Coulomb interactions on the dimer units are shown in 
 Table S3 in the Supplemental Material. 
We plot the pressure dependence of the
hopping integrals in Fig. \ref{fig-params} (c).  
In the H-phase, we find that $t_{11}$ and $t_{10}$ are the largest hopping integrals
and that their amplitudes are almost the same. 
Amplitudes other hopping integrals---such as $t_{01}$ and $t_{1m1}$---
are about 10\% of $t_{11}$ and $t_{10}$. 
This result shows that the low-energy effective Hamiltonians
for the H-phase 
is described 
as a nearly perfect rectangular lattice.
In all pressure ranges, deviations from a perfectly rectangular lattice is less than 10 \%.
This result is consistent with the fact that this compound has
a higher N\'eel temperature than other organic compounds.
Since geometrical frustration is absent, we expect the simple AF order shown in Fig. \ref{fig-params}(b) to be realized.

Fig. \ref{fig-params} (d) shows the pressure dependence of the normalized on-site Coulomb interaction $U/t_{11}$, which characterizes the amplitude of the electron correlations.
By increasing the pressure, we find that $U/t_{11}$ decreases almost monotonically.
This tendency is consistent with the decrease in the N\'eel temperatures under pressure.
This result indicates that the amplitudes of the electron correlations 
govern the insulator--superconductor transition under pressure. 

In Fig. \ref{fig-params} (e), we plot the normalized superexchange interactions $J_{ij}/J_{11}$.
We find that $J_{10}/J_{11}\sim 0.6$ for all pressure regions, while the other superexchange interactions
are almost zero. 
This result demonstrates that a nearly perfect rectangular lattice, without geometrical frustration, is realized in this compound.
Although square and rectangular lattices without geometrical frustration have been calculated theoretically, there are few reports of their existence in real materials.
Thus, (DMET--TTF)$_2$AuBr$_2$ offers an ideal platform for 
examining the relationship between bandwidth control of the
metal-insulator transition and the emergence of superconductivity.
Since the optimal superconducting transition temperature
$T_{\rm c}^{\rm opt}=4$K is not as high as that of
$\beta^{\prime}$-(BEDT--TTF)$_{2}$ICl$_{2}$ ($T_{\rm c}^{\rm opt}=14.2$K),
this result indicates that geometrical frustration may play an important role 
in increasing the superconducting transition temperature by
suppressing competing magnetic orders.

{\it Conclusions---.}From the x-ray diffraction, the resistivity measurement, and the $^{13}$C NMR measurements, we find that the organic conductor (DMET--TTF)$_2$AuBr$_2$ undergoes a pressure-induced structural phase transition at a pressure of around 0.9 GPa from a Q1D electronic system into a Q2D state. 
From $ab$ $initio$ calculations, we find that a nearly perfect rectangular lattice 
emerges in the H-phase.
This result is consistent
with the fact that this compound has a high N\'eel temperature.
Our study paves the way to create a 
unique lattice structure of molecular crystals
using pressure, which is not stabilized by other methods such as chemical substitution.

\begin{acknowledgments}
A part of this work was performed using facilities of the Institute for Solid State Physics, the University of Tokyo.
We would like to thank Mr. Yoshiya Sugawara and Ms. Kyoko Koyanagawa for their experimental supports.
This work was supported by the Japan Society for the Promotion of Science KAKENHI (Grant No. JP19H00648, JP19K0370709, JP20K14401, {21H01041, 22K03526}), and Hokkaido University (JP), Global Facility Center, Advanced Physical Property Open Unit, funded by MEXT (JP) under "Support Program for Implementation of New Equipment Sharing System" (JPMXS0420100318). 
\end{acknowledgments}
\newpage

{\Large
Supplemental Material for ``Pressure-induced nearly perfect rectangular lattice and superconductivity in an organic molecular crystal (DMET-TTF)$_2$AuBr$_2$"}

\section{1.~Synthesis and x-ray diffraction experiments}
We synthesized single crystals of (DMET-TTF)$_2$AuBr$_2$ using a standard electrochemical oxidation method. 
Each single crystal had the approximate shape of a parallelepiped. 
The typical size of 
a single crystal was 0.5 $\times$ 0.3 $\times$ 0.08 mm$^3$. 
The shortest side of the parallelepiped was the $c$-axis, and each of the longer sides of the parallelepiped was aligned with either the $a$- or $b$-axis. 
We investigated the properties of the synthesized single crystals using x-ray diffraction measurements obtained with a HyPix-6000 using Mo K-$\alpha$ radiation.
We established the crystal structure using ShelXT \cite{XT} and Olex2 \cite{olex2} and refined it using ShelXL \cite{XL}. 
We applied pressures up to 7.6 GPa using a DAC with a Re gasket.
The pressure was monitored with the ruby-fluorescence method. 
We used a 4:1 methanol--ethanol mixture as the pressure medium; it solidifies at 10 GPa at room temperature \cite{41}.

\section{2.~Resistivity measurements under pressure}
For the resistivity measurements along the $a$- and $b$-axes, we used the standard four-terminal method with carbon paste and gold wires. 
For the $c$-axis resistivity, we measured the inter-plane voltage against the inter-plane current by attaching terminals to both the front and back of the sample.
We applied the pressure using a hybrid NiCrAl-BeCu clamp cell, with Daphne 7373 oil \cite{7373} as the pressure medium.
We applied the current using a Keithley 220 direct-current source and measured the voltage using an Agilent 34420A nanovolt/microhm meter. 
We measured the pressure dependence of the resistivity along the $a$-, $b$-, and $c$-axes at room temperature at pressures up to 2.1 GPa as shown in Fig. 2(a) of the main paper. 

As shown in Fig. \ref{Anisotropy}(a), we found the in-plane anisotropy of the resistivity, $\rho_a$/$\rho_b$, to decrease monotonically with increasing pressures up to 0.8 GPa. At a pressure of around 0.9 GPa, it then decreases discontinuously from $\rho_a$/$\rho_b = 18$ to $5.6$.
It subsequently decreases monotonically again as the pressure increases up to 2.1 GPa. 
Thus, a large decrease in the anisotropy of the resistivity occurs at the aforementioned resistivity discontinuity. 
Since the conductivity of a material is proportional to the square of the transfer integral, $t$, the in-plane anisotropy of the transfer integral, $t_b$/$t_a$, thus decreases discontinuously from 4.2 to 2.4 at a pressure of around 0.9 GPa. 
This implies that the electronic state undergoes a transition from a Q1D electronic state to a quasi-two-dimensional (Q2D) state at the resistance discontinuity.

We also measured the temperature dependence of the resistivity along the $c$-axis under pressures of 2.5--15.0 GPa using a constant-loading-type CAC \cite{CAC} 
and under 5.0--9.0 GPa (at low temperatures) using a clamp-type PCAC \cite{PCAC}.
We used Daphne 7373 oil as the pressure medium. 
We measured the pressure dependence of the resistivity along the $c$-axis at room temperature on sample 8 with applied pressures up to 2.0 GPa 
using a CAC with Fluorinert (FC-70:FC-77 = 1:1) as the pressure medium. 
Since the pressure is applied using a cubic six-anvil cell, it is almost equivalent to hydrostatic pressure. 
The pressure in the CAC remained almost independent of the temperature during the slow heating process, whereas the pressure of the PCAC increased with decreasing temperature \cite{PCAC}
(e.g., 4.1 GPa at room temperature becomes 6.0 GPa at 7 K). 
For the resistivity measurements using the PCAC, we therefore used only those measurements taken at temperatures below 20 K.
We applied the current using a Keithley 2400 direct-current source, and we measured the voltage using a Keithley 2182 nanovolt meter. 
We used a dilution refrigerator to perform measurements at temperatures below 6 K and magnetic fields up to 3 T.
We also measured the magnetic-field dependence of the resistivity up to fields of 3 T at temperatures of 6.4 K, 2.0 K, and 26 mK.
We performed AC electrical-resistance measurements using a Linear Research LR-700 AC resistance bridge;
the current used for these measurements was 1 $\rm \mu$A.
We applied the magnetic field normal to the conducting plane. 

The details of the setup for the cubic anvil cell are as follows: 
After attaching four gold wires to the sample using carbon paste, we placed the sample in a Teflon cell, which we then filled with a pressure medium. 
We then placed a cubic MgO gasket containing the Teflon cell in the center of the six-anvil cell \cite{ICl2_Taniguchi,AuCl2_Taniguchi}.

\begin{figure*}[t]
\centering 
\includegraphics[width=160mm]{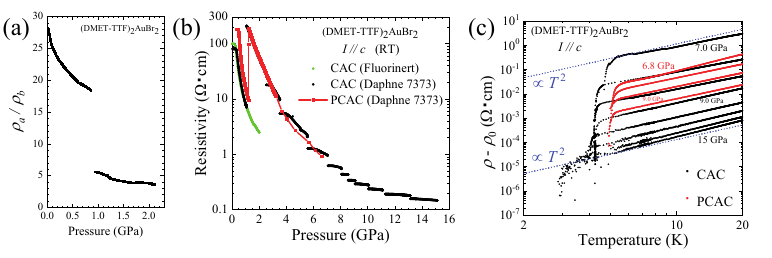}
\caption{
(a) Pressure dependence of the in-plane anisotropy of the resistivity, $\rho_a$/$\rho_b$, at room temperature with pressures up to 2.1 GPa.
(b) Pressure dependence of the $c$-axis resistivity at room temperature using a CAC and a PCAC with Daphne 7373 oil, as well as a CAC with Fluorinert. 
(c) Temperature dependence of the $c$-axis resistivity below 20 K for pressures in the range 7.0--15 GPa (CAC) and 6.8--9.0 GPa (PCAC). 
Residual resistivity $\rho_0$ is subtracted from the plotted values. The blue dotted lines are proportional to $T^2$. 
}
\label{Anisotropy}
\end{figure*}

Figure \ref{Anisotropy}(b) shows the pressure dependence of the $c$-axis resistivity for pressures up to 15 GPa. The structural phase transition was confirmed to occur at around 1.0 GPa when Daphne 7373 oil was used as the pressure medium. 
The $c$-axis resistivity at 15 GPa was reduced to 1/1000 of that observed at 1.0 GPa. 
However, when Fluorinert (FC-70:FC-77 = 1:1) was used as the pressure medium, we observed no 
resistance jump for pressures up to 2.0 GPa, even when hydrostatic pressure was applied using the CAC. 
The solidification pressure of
the Fluorinert (FC-70:FC-77 = 1:1) is $\sim$0.85 GPa \cite{Fluorinert}.
In contrast, Daphne 7373 oil solidifies at 2.2 GPa at room temperature \cite{Daphne7373}.
This suggests that the structural phase transition is accompanied by a crystal deformation that is suppressed by the solidification of the pressure medium.

As shown in Fig. \ref{Anisotropy}(c), the metallic phase above the SC transition temperature follows the power law $T^{\alpha} (\alpha = 1.75$--$2.28)$ between 6.8 GPa and 15 GPa for temperatures below 20 K, which indicates Fermi-liquid-like behavior. 
We thus expect (DMET--TTF)$_2$AuBr$_2$ to have a Q2D Fermi surface, 
with a Fermi-liquid-like metallic phase existing just above $T_{\rm c}$.

\section{3.~$^{13}$C--NMR measurements under pressure and the antiferromagnetic transition}
We synthesized single-site $^{13}$C-substituted DMET--TTF 
molecules via the cross-coupling method 
for $^{13}$C-NMR measurements~\cite{coupling_1,coupling_2}
We fixed the conducting plane to the inside of a coil with Araldite.
We then placed the coil in the hybrid NiCrAl-BeCu clamp cell, and we applied pressure using Daphne 7373 oil. 
We applied a 6T magnetic field normal to the conducting plane, which we confirmed by measuring the chemical shift and hyperfine coupling constant at ambient pressure and comparing them with previous data \cite{AuBr2_AP_C-SDW}. 
We obtained the $^{13}$C-NMR spectra via fast Fourier transformations of the echo signals from $\pi/2-\pi$ pulse sequences. 
A typical value of the $\pi/2$ pulse length was 1.5 $\mu$s.
We referenced the $^{13}$C-NMR shifts to that of tetramethylsilane (TMS). 
We measured the spin-lattice relaxation time, $T_1$, using the conventional saturation--recovery method.

%FIGURE AFM%
\begin{figure*}
\centering
\includegraphics[width=120mm]{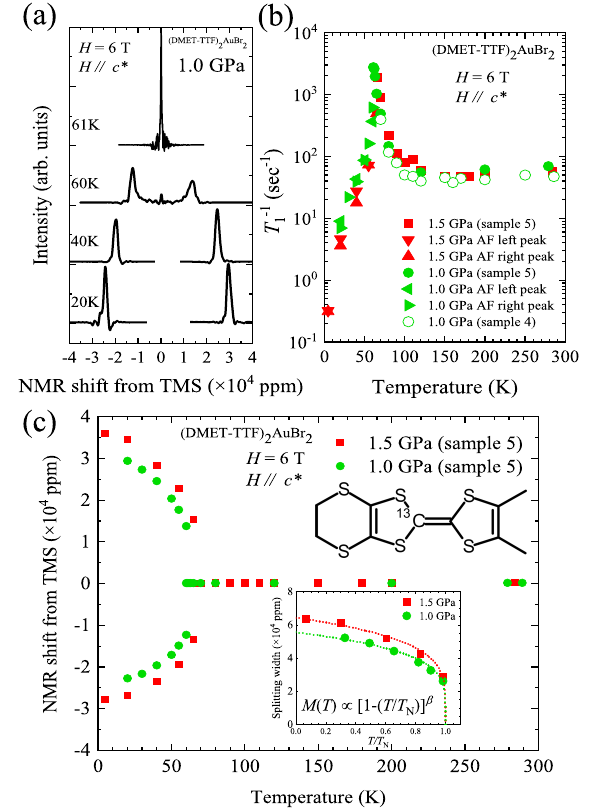}
\caption{
(a) $^{13}$C-NMR spectra of (DMET--TTF)$_2$AuBr$_2$ below 61 K at a pressure of 1.0 GPa. 
(b) The spin-lattice relaxation rate $T_1^{-1}$ exhibits a peak structure at 61 K at a pressure of 1.0 GPa and at 66 K under a pressure of 1.5 GPa. 
(c) $^{13}$C-NMR shift at 1.0 and 1.5 GPa. 
The inset shows the splitting widths at pressures of 1.0 and 1.5 GPa. 
The critical exponent, $\beta$ (see the formula in the inset), is equal to 0.21 at both pressures. 
We performed these measurements on DMET--TTF molecules with $^{13}$C substituted on one side of the double bond.
}
\label{AFM}
\end{figure*}
In previous work, the hyperfine coupling tensor of (DMET--TTF)$_2$AuBr$_2$ was determined at ambient pressure using $^{13}$C-NMR measurements, 
and the magnetic moment of the C-SDW was found to be 0.12 $\mu_{\rm B}$/dimer at 14 K \cite{AuBr2_AP_C-SDW}.
The $^{13}$C-NMR spectrum of (DMET--TTF)$_2$AuBr$_2$ in the paramagnetic phase at ambient pressure is characterized by a single peak because there is only one crystallographically equivalent $^{13}$C in each unit cell of the L-phase. 

To investigate the electronic state of the H-phase, 
we obtained $^{13}$C-NMR spectra at 65.11 MHz under pressures of 1.0 and 1.5 GPa. 
Figure \ref{AFM}(a) shows the NMR spectra at a pressure of 1.0 GPa. 
Only one peak is observed in the spectra above 61 K.  
This result demonstrates that there is only
one crystallographically equivalent $^{13}$C site in each unit cell of the H-phase. 
However, as shown in Fig. \ref{AFM}(a), 
the single NMR peak splits below 61 K at a pressure of 1.0 GPa and below 66 K at a pressure of 1.5 GPa. 
Figure \ref{AFM}(b) shows the spin-lattice relaxation rate, $T_1^{-1}$.
It diverges due to critical slowing down at 61 K at 1.0 GPa and at 66 K at 1.5 GPa; below the transition temperature,
the relaxation rate decreases rapidly with decreasing temperature. 
This behavior provides strong evidence of an AF transition. 
The AF transition temperature of (DMET--TTF)$_2$AuBr$_2$ is $T_{\rm N} = 61$ K at a pressure of 1.0 GPa and 66 K at a pressure of 1.5 GPa; 
These AF transition temperatures are rather high AF transition temperatures for a low-dimensional organic conductor when compared with 27 K 
for $\mathrm{\kappa}$-(BEDT--TTF)$_2$Cu[N(CN)$_2$]Cl \cite{kET_AF} 
and 22 K for $\mathrm{\beta}'$-(BEDT--TTF)$_2$ICl$_2$ \cite{ICl2_TN}
at ambient pressure.

The magnetic-field direction is the same as that employed in previous $^{13}$C-NMR measurements at ambient pressure \cite{AuBr2_AP_C-SDW}, 
where the magnetic moment of the commensurate spin-density-wave (C-SDW) of (DMET-TTF)$_2$AuBr$_2$ was estimated to be 0.12 $\mu_{\rm B}$/dimer at ambient pressure.
The splitting widths of the NMR spectra at the pressures 1.0 and 1.5 GPa are more than 30 times larger than that observed at ambient pressure in the C-SDW state.
{Although the exact determination of the hyperfine coupling tensor of the H phase was experimentally difficult,}
the AF phase of (DMET--TTF)$_2$AuBr$_2$ in the H-phase appears to be the typical AF state of  a dimer Mott insulator, with a magnetic moment of 1 $\mu_{\rm B}$/dimer.

Figure \ref{AFM} (c) shows that the splitting width of the NMR spectrum at a pressure of 1.5 GPa is slightly larger than that observed at 1.0 GPa. 
The magnetic moment in the AF phase can be fitted with an exponent formula {as $M(T)\propto \left[1-(T/T_{\rm N})\right]^\beta$}[see the inset in Fig. \ref{AFM}(c)]. {We find
$\beta=0.21$ for both pressures which are close to the sub-critical exponent of a two-dimensional $XY$ model, where $\beta=0.23$ \cite{2DXY}.}
Above $T_{\rm N}$, $T_1^{-1}$ is constant [see Fig. \ref{AFM}(b)], which suggests that localized spins exist in this electronic system. 
The paramagnetic insulating phase above $T_{\rm N}$ is, therefore, likely to be a dimer Mott insulator phase.

\section{4.~Derivation of the ab initio effective Hamiltonians}
We performed 
{\it ab initio} calculations using \verb|Quantum ESPRESSO| (version 6.8) \cite{QE}.
We employed norm-conserving pseudopotentials based on the Vanderbilt formalism with plane-wave basis sets~\cite{Hamann_ONCV2013, Schlipf_CPC2015}. 
As the exchange-correlation functional, we used the generalized gradient approximation by Perdew, Burke, and Ernzerhof~\cite{GGA_PBE}.
The cutoff energies for the plane waves and charge densities were $70$ Ry and $280$ Ry, respectively.
We used a $5\times 5\times 3$ uniform $\bm{k}$-point mesh with a Gaussian smearing method during self-consistent loops. 
We relaxed the positions of all the atoms using 
{\it ab initio} calculations based on experimental crystal structures.

%TABLE1%
\begin{table*}[h]
\caption{
The lattice parameters of (DMET--TTF)$_2$AuBr$_2$ at applied pressures of 0, 0.9, and 1.7 GPa.
}
\begin{tabular}{l c c c c c c c}
\hline\hline
(DMET-TTF)$_2$AuBr$_2$ & $a$ (\AA) & $b$ (\AA) & $c$ (\AA) & $\alpha$ ($^\circ$) & $\beta$ ($^\circ$) & $\gamma$ ($^\circ$)	& $V$ (\AA$^3$)	\\
\hline
0 GPa & 6.6885(10) & 7.6737(16) & 15.490(5) & 91.41(3) & 101.51(3) &  103.325(16) & 756.0(3)  \\
0.9 GPa	& 6.4561(10) & 7.3590(13) & 15.151(5) & 89.49 (3) & 100.42(3) & 103.086(14) & 689.2(3)  \\
1.7 GPa	& 6.390(2) & 7.043(2) & 16.07(4) & 104.6(2) & 99.2(3) & 104.68(3) & 657(2)  \\
\hline\hline
\end{tabular}
\label{lattice}
\end{table*}

\begin{figure*}
    \centering
    \includegraphics[width=160mm]{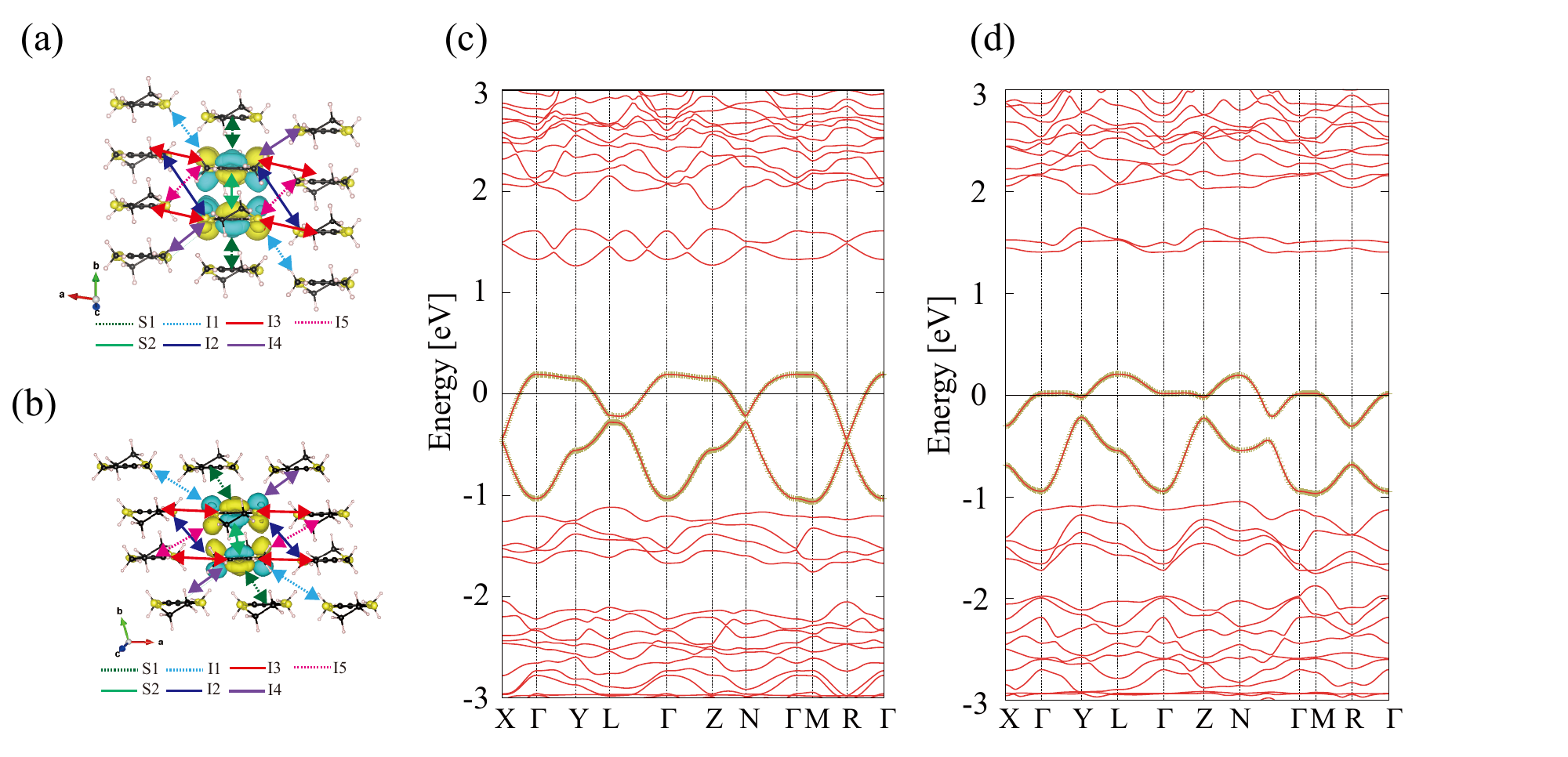}
    \caption{The maximally localized Wannier functions (MLFs) in the dimer picture for (DMET--TTF)$_2$AuBr$_2$ at the pressure $P$ = (a) $0.9$ GPa (L--phase) and (b) $1.7$ GPa (H--phase). The arrows indicate bonding between the MLWFs. Band structure of (DMET--TTF)$_2$AuBr$_2$ at a pressure of (c) $0.9$ GPa and (d) $1.7$ GPa. The thin red solid lines are density-functional theory (DFT) results. The bold green lines are obtained by using maximally localized Wannier functions (MLWFs). We set the Fermi energy to zero (the solid line). Here, we define following indices: X=$(0, -\pi, 0)$, $\Gamma=(0,0, 0)$, Y=$(\pi, 0, 0)$, L=$(\pi, -\pi, 0)$, Z=$(-\pi, 0, \pi)$, N=$(-\pi, -\pi, \pi)$, M=$(0, 0, \pi)$, and R=$(0, -\pi, \pi)$. 
    }
    \label{fig:bands}
\end{figure*}
%TABLE 2%
\begin{table*}[h]
\caption{
The transfer integrals of (DMET--TTF)$_2$AuBr$_2$ at the pressures $0.9$ and $1.7$ GPa were obtained using maximally localized Wannier functions. The energy units are eV.}
\label{overlap}
\centering
\begin{tabular}{l c c c c c c c c} 
\hline\hline
(DMET-TTF)$_2$AuBr$_2$ & $t_{\mathrm{S1}}$ & $t_{\mathrm{S2}}$ & $t_{\mathrm{I1}}$ & $t_{\mathrm{I2}}$ & $t_{\mathrm{I3}}$ &  $t_{\mathrm{I4}}$ & $t_{\mathrm{I5}}$ \\
\hline
0.9 GPa	& -0.256 & -0.240 & 0.005 & 0.005 & -0.056 & -0.066 & -0.079 \\
1.7 GPa	& -0.003 & -0.289 & 0.015 & -0.005 & -0.085& -0.157 & -0.047 \\ \hline\hline
\end{tabular}
\label{table:transfer}
\end{table*}

Using the electronic states obtained from the
{\it ab initio} calculations, we constructed maximally localized Wannier functions (MLWFs) at $P= 0.9$ (L-phase) and $1.7$ GPa (H-phase) as shown in Figs. \ref{fig:bands}(a) and (b). We also show the band structures obtained by 
the {\it ab initio} calculations for (DMET--TTF)$_2$AuBr$_2$ at 0.9 GPa and 1.7 GPa are shown in Figs. \ref{fig:bands} (c) and (d), respectively. 
Here, the thin red solid lines are density-functional theory (DFT) results and the bold green lines are obtained by using MLWFs.
The values of transfer integrals are listed in TABLE \ref{table:transfer}. We can see that $|t_{S1}|$ is about the same size as $|t_{S2}|$ at $0.9$ GPa (L-phase), whereas $|t_{S2}|$ is larger than $|t_{S2}|$ at $1.7$ GPa (H-phase), indicating that dimerization occurs in the H-phase. 

Focusing on the stabilization of the dimer structure in the H-phase, we constructed one MLWF for one band crossing the Fermi energy as shown in Fig. 4(a) of the main paper.
Using the MLWF, we derived the interaction parameters in the H-phase using the constrained random phase approximation (cRPA)~\cite{PhysRevB.70.195104, Imada_JPSJ2010}. The obtained transfer and interaction parameters are listed in TABLE \ref{overlap2}.
Note that in many molecular conductors, the energy bands near the Fermi surface are isolated from the other bands.
This feature enables the construction of effective models with good accuracy. In fact, the cRPA method has been applied to several organic materials and has succeeded in reproducing abundant phenomena ~\cite{Shinaoka2012, PhysRevResearch.2.032072, PhysRevResearch.3.043224, Ido2022, PhysRevB.105.205123, PhysRevB.102.235116, yoshimi2022}.
To construct the MLWFs and derive the interaction parameters, we used the software package \verb|RESPACK|~\cite{RESPACK}, with the energy cutoff for the dielectric function set at $3$ Ry.

\begin{table*}[h]
\caption{
Transfer integrals and Coulomb interactions on the dimer units in the H-phase of (DMET-TTF)$_2$AuBr$_2$. The energy units are eV.}
\label{overlap2}
\centering
\begin{tabular}{l c c c c c c c c c} 
\hline\hline
P [GPa] & $t_{11}$ & $t_{10}$ & $t_{01}$ & $t_{1m1}$ & $U$ & $V_{11}$ & $V_{10}$ & $V_{01}$ & $V_{1m1}$\\
\hline
0.55  &  0.066 & -0.054 & 0.004 & -0.004 & 1.037 & 0.416 & 0.396 & 0.396 & 0.266\\
1.46 & 0.072 & -0.053 & 0.003 & -0.007 & 0.936 & 0.380 & 0.365 & 0.364 & 0.260 \\
1.7 & 0.074 & -0.059 & 0.005 & -0.007 & 0.967 & 0.392 & 0.376 & 0.376 & 0.259\\
2.04 & 0.075 & -0.059 & 0.003 & -0.007 & 0.948 & 0.390 & 0.371 & 0.372 & 0.258\\
2.825 & 0.078 & -0.065 & 0.002 & -0.009 & 0.917 & 0.373 & 0.360 & 0.362 & 0.251\\
3.17 & 0.080 & -0.066 & 0.000 & -0.009 & 0.902 & 0.366 & 0.355 & 0.356 & 0.247\\
4.085 & 0.082 & -0.070 & 0.000 & -0.010 & 0.880 & 0.352 & 0.345 & 0.348 & 0.242\\
5.14 & 0.085 & -0.073 & 0.001 & -0.012 & 0.845 & 0.335 & 0.334 & 0.336 & 0.236\\
5.73 & 0.086 & -0.071 & 0.002 & -0.011 & 0.828 & 0.326 & 0.330 & 0.330 & 0.238\\
6.51 & 0.088 & -0.077 & 0.000 & -0.014 & 0.815 & 0.320 & 0.323 & 0.326 & 0.229\\
7.61 & 0.091 & -0.074 & 0.010 & -0.014 & 0.827 & 0.330 & 0.331 & 0.334 & 0.235 \\ \hline\hline
\end{tabular}
\end{table*}

\clearpage
 %TC:endignore
%REFERENCE%
 %TC:ignore
\bibliography{main}

\providecommand{\noopsort}[1]{}\providecommand{\singleletter}[1]{#1}%
\begin{thebibliography}{57}%
\makeatletter
\providecommand \@ifxundefined [1]{%
 \@ifx{#1\undefined}
}%
\providecommand \@ifnum [1]{%
 \ifnum #1\expandafter \@firstoftwo
 \else \expandafter \@secondoftwo
 \fi
}%
\providecommand \@ifx [1]{%
 \ifx #1\expandafter \@firstoftwo
 \else \expandafter \@secondoftwo
 \fi
}%
\providecommand \natexlab [1]{#1}%
\providecommand \enquote  [1]{``#1''}%
\providecommand \bibnamefont  [1]{#1}%
\providecommand \bibfnamefont [1]{#1}%
\providecommand \citenamefont [1]{#1}%
\providecommand \href@noop [0]{\@secondoftwo}%
\providecommand \href [0]{\begingroup \@sanitize@url \@href}%
\providecommand \@href[1]{\@@startlink{#1}\@@href}%
\providecommand \@@href[1]{\endgroup#1\@@endlink}%
\providecommand \@sanitize@url [0]{\catcode `\\12\catcode `\$12\catcode
  `\&12\catcode `\#12\catcode `\^12\catcode `\_12\catcode `\%12\relax}%
\providecommand \@@startlink[1]{}%
\providecommand \@@endlink[0]{}%
\providecommand \url  [0]{\begingroup\@sanitize@url \@url }%
\providecommand \@url [1]{\endgroup\@href {#1}{\urlprefix }}%
\providecommand \urlprefix  [0]{URL }%
\providecommand \Eprint [0]{\href }%
\providecommand \doibase [0]{https://doi.org/}%
\providecommand \selectlanguage [0]{\@gobble}%
\providecommand \bibinfo  [0]{\@secondoftwo}%
\providecommand \bibfield  [0]{\@secondoftwo}%
\providecommand \translation [1]{[#1]}%
\providecommand \BibitemOpen [0]{}%
\providecommand \bibitemStop [0]{}%
\providecommand \bibitemNoStop [0]{.\EOS\space}%
\providecommand \EOS [0]{\spacefactor3000\relax}%
\providecommand \BibitemShut  [1]{\csname bibitem#1\endcsname}%
\let\auto@bib@innerbib\@empty
%</preamble>
\bibitem [{\citenamefont {Gao}\ \emph {et~al.}(1994)\citenamefont {Gao},
  \citenamefont {Xue}, \citenamefont {Chen}, \citenamefont {Xiong},
  \citenamefont {Meng}, \citenamefont {Ramirez}, \citenamefont {Chu},
  \citenamefont {Eggert},\ and\ \citenamefont {Mao}}]{Gao_PRB1994}%
  \BibitemOpen
  \bibfield  {author} {\bibinfo {author} {\bibfnamefont {L.}~\bibnamefont
  {Gao}}, \bibinfo {author} {\bibfnamefont {Y.~Y.}\ \bibnamefont {Xue}},
  \bibinfo {author} {\bibfnamefont {F.}~\bibnamefont {Chen}}, \bibinfo {author}
  {\bibfnamefont {Q.}~\bibnamefont {Xiong}}, \bibinfo {author} {\bibfnamefont
  {R.~L.}\ \bibnamefont {Meng}}, \bibinfo {author} {\bibfnamefont
  {D.}~\bibnamefont {Ramirez}}, \bibinfo {author} {\bibfnamefont {C.~W.}\
  \bibnamefont {Chu}}, \bibinfo {author} {\bibfnamefont {J.~H.}\ \bibnamefont
  {Eggert}},\ and\ \bibinfo {author} {\bibfnamefont {H.~K.}\ \bibnamefont
  {Mao}},\ }\bibfield  {title} {\bibinfo {title} {{Superconductivity up to 164
  K in HgBa$_2$Ca$_{m-1}$Cu$_{m}$O$_{2m+2+\delta}$ ($m$=1, 2, and 3) under
  quasihydrostatic pressures}},\ }\href
  {https://doi.org/10.1103/PhysRevB.50.4260} {\bibfield  {journal} {\bibinfo
  {journal} {Phys. Rev. B}\ }\textbf {\bibinfo {volume} {50}},\ \bibinfo
  {pages} {4260} (\bibinfo {year} {1994})}\BibitemShut {NoStop}%
\bibitem [{\citenamefont {Monteverde}\ \emph {et~al.}(2005)\citenamefont
  {Monteverde}, \citenamefont {Acha}, \citenamefont {NC:C1ez-Regueiro},
  \citenamefont {Pavlov}, \citenamefont {Lokshin}, \citenamefont {Putilin},\
  and\ \citenamefont {Antipov}}]{Monteverde_EPL2005}%
  \BibitemOpen
  \bibfield  {author} {\bibinfo {author} {\bibfnamefont {M.}~\bibnamefont
  {Monteverde}}, \bibinfo {author} {\bibfnamefont {C.}~\bibnamefont {Acha}},
  \bibinfo {author} {\bibfnamefont {M.}~\bibnamefont {NC:C1ez-Regueiro}},
  \bibinfo {author} {\bibfnamefont {D.~A.}\ \bibnamefont {Pavlov}}, \bibinfo
  {author} {\bibfnamefont {K.~A.}\ \bibnamefont {Lokshin}}, \bibinfo {author}
  {\bibfnamefont {S.~N.}\ \bibnamefont {Putilin}},\ and\ \bibinfo {author}
  {\bibfnamefont {E.~V.}\ \bibnamefont {Antipov}},\ }\bibfield  {title}
  {\bibinfo {title} {{High-pressure effects in fluorinated
  HgBa$_2$Ca$_2$Cu$_3$O$_{8+\delta}$}},\ }\href
  {https://doi.org/10.1209/epl/i2005-10247-3} {\bibfield  {journal} {\bibinfo
  {journal} {Europhysics Letters}\ }\textbf {\bibinfo {volume} {72}},\ \bibinfo
  {pages} {458} (\bibinfo {year} {2005})}\BibitemShut {NoStop}%
\bibitem [{\citenamefont {Takeshita}\ \emph {et~al.}(2013)\citenamefont
  {Takeshita}, \citenamefont {Yamamoto}, \citenamefont {Iyo},\ and\
  \citenamefont {Eisaki}}]{Takeshita_JPSJ2013}%
  \BibitemOpen
  \bibfield  {author} {\bibinfo {author} {\bibfnamefont {N.}~\bibnamefont
  {Takeshita}}, \bibinfo {author} {\bibfnamefont {A.}~\bibnamefont {Yamamoto}},
  \bibinfo {author} {\bibfnamefont {A.}~\bibnamefont {Iyo}},\ and\ \bibinfo
  {author} {\bibfnamefont {H.}~\bibnamefont {Eisaki}},\ }\bibfield  {title}
  {\bibinfo {title} {{Zero Resistivity above 150 K in
  HgBa$_2$Ca$_2$Cu$_3$O$_{8+\delta}$ at High Pressure}},\ }\href
  {https://doi.org/10.7566/JPSJ.82.023711} {\bibfield  {journal} {\bibinfo
  {journal} {J. Phys. Soc. Jpn.}\ }\textbf {\bibinfo {volume} {82}},\ \bibinfo
  {pages} {023711} (\bibinfo {year} {2013})}\BibitemShut {NoStop}%
\bibitem [{\citenamefont {Takahashi}\ \emph {et~al.}(2008)\citenamefont
  {Takahashi}, \citenamefont {Igawa}, \citenamefont {Arii}, \citenamefont
  {Kamihara}, \citenamefont {Hirano},\ and\ \citenamefont
  {Hosono}}]{Takahashi_Nature2008}%
  \BibitemOpen
  \bibfield  {author} {\bibinfo {author} {\bibfnamefont {H.}~\bibnamefont
  {Takahashi}}, \bibinfo {author} {\bibfnamefont {K.}~\bibnamefont {Igawa}},
  \bibinfo {author} {\bibfnamefont {K.}~\bibnamefont {Arii}}, \bibinfo {author}
  {\bibfnamefont {Y.}~\bibnamefont {Kamihara}}, \bibinfo {author}
  {\bibfnamefont {M.}~\bibnamefont {Hirano}},\ and\ \bibinfo {author}
  {\bibfnamefont {H.}~\bibnamefont {Hosono}},\ }\bibfield  {title} {\bibinfo
  {title} {{Superconductivity at 43 K in an iron-based layered compound
  LaO$_{1-x}$F$_x$FeAs}},\ }\href
  {https://doi.org/https://doi.org/10.1038/nature06972} {\bibfield  {journal}
  {\bibinfo  {journal} {Nature}\ }\textbf {\bibinfo {volume} {453}},\ \bibinfo
  {pages} {376} (\bibinfo {year} {2008})}\BibitemShut {NoStop}%
\bibitem [{\citenamefont {Stewart}(2011)}]{Stewart_RMP2011}%
  \BibitemOpen
  \bibfield  {author} {\bibinfo {author} {\bibfnamefont {G.~R.}\ \bibnamefont
  {Stewart}},\ }\bibfield  {title} {\bibinfo {title} {Superconductivity in iron
  compounds},\ }\href {https://doi.org/10.1103/RevModPhys.83.1589} {\bibfield
  {journal} {\bibinfo  {journal} {Rev. Mod. Phys.}\ }\textbf {\bibinfo {volume}
  {83}},\ \bibinfo {pages} {1589} (\bibinfo {year} {2011})}\BibitemShut
  {NoStop}%
\bibitem [{\citenamefont {Wang}\ \emph {et~al.}(2022)\citenamefont {Wang},
  \citenamefont {Yang}, \citenamefont {Yang}, \citenamefont {Chen},
  \citenamefont {Zhang}, \citenamefont {Zhang}, \citenamefont {Zhu},
  \citenamefont {Uwatoko}, \citenamefont {Gu}, \citenamefont {Dong} \emph
  {et~al.}}]{Wang_NCom2022}%
  \BibitemOpen
  \bibfield  {author} {\bibinfo {author} {\bibfnamefont {N.}~\bibnamefont
  {Wang}}, \bibinfo {author} {\bibfnamefont {M.}~\bibnamefont {Yang}}, \bibinfo
  {author} {\bibfnamefont {Z.}~\bibnamefont {Yang}}, \bibinfo {author}
  {\bibfnamefont {K.}~\bibnamefont {Chen}}, \bibinfo {author} {\bibfnamefont
  {H.}~\bibnamefont {Zhang}}, \bibinfo {author} {\bibfnamefont
  {Q.}~\bibnamefont {Zhang}}, \bibinfo {author} {\bibfnamefont
  {Z.}~\bibnamefont {Zhu}}, \bibinfo {author} {\bibfnamefont {Y.}~\bibnamefont
  {Uwatoko}}, \bibinfo {author} {\bibfnamefont {L.}~\bibnamefont {Gu}},
  \bibinfo {author} {\bibfnamefont {X.}~\bibnamefont {Dong}}, \emph {et~al.},\
  }\bibfield  {title} {\bibinfo {title} {{Pressure-induced monotonic
  enhancement of T c to over 30 K in superconducting Pr0. 82Sr0. 18NiO2 thin
  films}},\ }\href {https://doi.org/https://doi.org/10.1038/s41467-022-32065-x}
  {\bibfield  {journal} {\bibinfo  {journal} {Nature communications}\ }\textbf
  {\bibinfo {volume} {13}},\ \bibinfo {pages} {4367} (\bibinfo {year}
  {2022})}\BibitemShut {NoStop}%
\bibitem [{\citenamefont {Sun}\ \emph {et~al.}(2023)\citenamefont {Sun},
  \citenamefont {Huo}, \citenamefont {Hu}, \citenamefont {Li}, \citenamefont
  {Liu}, \citenamefont {Han}, \citenamefont {Tang}, \citenamefont {Mao},
  \citenamefont {Yang}, \citenamefont {Wang} \emph {et~al.}}]{Sun_Nature2023}%
  \BibitemOpen
  \bibfield  {author} {\bibinfo {author} {\bibfnamefont {H.}~\bibnamefont
  {Sun}}, \bibinfo {author} {\bibfnamefont {M.}~\bibnamefont {Huo}}, \bibinfo
  {author} {\bibfnamefont {X.}~\bibnamefont {Hu}}, \bibinfo {author}
  {\bibfnamefont {J.}~\bibnamefont {Li}}, \bibinfo {author} {\bibfnamefont
  {Z.}~\bibnamefont {Liu}}, \bibinfo {author} {\bibfnamefont {Y.}~\bibnamefont
  {Han}}, \bibinfo {author} {\bibfnamefont {L.}~\bibnamefont {Tang}}, \bibinfo
  {author} {\bibfnamefont {Z.}~\bibnamefont {Mao}}, \bibinfo {author}
  {\bibfnamefont {P.}~\bibnamefont {Yang}}, \bibinfo {author} {\bibfnamefont
  {B.}~\bibnamefont {Wang}}, \emph {et~al.},\ }\bibfield  {title} {\bibinfo
  {title} {{Signatures of superconductivity near 80 K in a nickelate under high
  pressure}},\ }\href
  {https://doi.org/https://doi.org/10.1038/s41586-023-06408-7} {\bibfield
  {journal} {\bibinfo  {journal} {Nature}\ }\textbf {\bibinfo {volume} {621}},\
  \bibinfo {pages} {493} (\bibinfo {year} {2023})}\BibitemShut {NoStop}%
\bibitem [{\citenamefont {Wang}\ \emph {et~al.}(2024)\citenamefont {Wang},
  \citenamefont {Wang}, \citenamefont {Shen}, \citenamefont {Hou},
  \citenamefont {Ma}, \citenamefont {Shi}, \citenamefont {Ren}, \citenamefont
  {Gu}, \citenamefont {Ma}, \citenamefont {Yang}, \citenamefont {Liu},
  \citenamefont {Guo}, \citenamefont {Sun}, \citenamefont {Zhang},
  \citenamefont {Calder}, \citenamefont {Yan}, \citenamefont {Wang},
  \citenamefont {Uwatoko},\ and\ \citenamefont {Cheng}}]{Wang_PRX2024}%
  \BibitemOpen
  \bibfield  {author} {\bibinfo {author} {\bibfnamefont {G.}~\bibnamefont
  {Wang}}, \bibinfo {author} {\bibfnamefont {N.~N.}\ \bibnamefont {Wang}},
  \bibinfo {author} {\bibfnamefont {X.~L.}\ \bibnamefont {Shen}}, \bibinfo
  {author} {\bibfnamefont {J.}~\bibnamefont {Hou}}, \bibinfo {author}
  {\bibfnamefont {L.}~\bibnamefont {Ma}}, \bibinfo {author} {\bibfnamefont
  {L.~F.}\ \bibnamefont {Shi}}, \bibinfo {author} {\bibfnamefont {Z.~A.}\
  \bibnamefont {Ren}}, \bibinfo {author} {\bibfnamefont {Y.~D.}\ \bibnamefont
  {Gu}}, \bibinfo {author} {\bibfnamefont {H.~M.}\ \bibnamefont {Ma}}, \bibinfo
  {author} {\bibfnamefont {P.~T.}\ \bibnamefont {Yang}}, \bibinfo {author}
  {\bibfnamefont {Z.~Y.}\ \bibnamefont {Liu}}, \bibinfo {author} {\bibfnamefont
  {H.~Z.}\ \bibnamefont {Guo}}, \bibinfo {author} {\bibfnamefont {J.~P.}\
  \bibnamefont {Sun}}, \bibinfo {author} {\bibfnamefont {G.~M.}\ \bibnamefont
  {Zhang}}, \bibinfo {author} {\bibfnamefont {S.}~\bibnamefont {Calder}},
  \bibinfo {author} {\bibfnamefont {J.-Q.}\ \bibnamefont {Yan}}, \bibinfo
  {author} {\bibfnamefont {B.~S.}\ \bibnamefont {Wang}}, \bibinfo {author}
  {\bibfnamefont {Y.}~\bibnamefont {Uwatoko}},\ and\ \bibinfo {author}
  {\bibfnamefont {J.-G.}\ \bibnamefont {Cheng}},\ }\bibfield  {title} {\bibinfo
  {title} {{Pressure-Induced Superconductivity In Polycrystalline
  ${\mathrm{La}}_{3}{\mathrm{Ni}}_{2}{\mathrm{O}}_{7\ensuremath{-}\ensuremath{\delta}}$}},\
  }\href {https://doi.org/10.1103/PhysRevX.14.011040} {\bibfield  {journal}
  {\bibinfo  {journal} {Phys. Rev. X}\ }\textbf {\bibinfo {volume} {14}},\
  \bibinfo {pages} {011040} (\bibinfo {year} {2024})}\BibitemShut {NoStop}%
\bibitem [{\citenamefont {Drozdov}\ \emph {et~al.}(2015)\citenamefont
  {Drozdov}, \citenamefont {Eremets}, \citenamefont {Troyan}, \citenamefont
  {Ksenofontov},\ and\ \citenamefont {Shylin}}]{SCH2S}%
  \BibitemOpen
  \bibfield  {author} {\bibinfo {author} {\bibfnamefont {A.~P.}\ \bibnamefont
  {Drozdov}}, \bibinfo {author} {\bibfnamefont {M.~I.}\ \bibnamefont
  {Eremets}}, \bibinfo {author} {\bibfnamefont {I.~A.}\ \bibnamefont {Troyan}},
  \bibinfo {author} {\bibfnamefont {V.}~\bibnamefont {Ksenofontov}},\ and\
  \bibinfo {author} {\bibfnamefont {S.~I.}\ \bibnamefont {Shylin}},\ }\bibfield
   {title} {\bibinfo {title} {{Conventional superconductivity at 203 kelvin at
  high pressures in the sulfur hydride system}},\ }\href
  {https://doi.org/10.1038/nature14964} {\bibfield  {journal} {\bibinfo
  {journal} {Nature}\ }\textbf {\bibinfo {volume} {525}},\ \bibinfo {pages}
  {73} (\bibinfo {year} {2015})}\BibitemShut {NoStop}%
\bibitem [{\citenamefont {Drozdov}\ \emph {et~al.}(2019)\citenamefont
  {Drozdov}, \citenamefont {Kong}, \citenamefont {Minkov}, \citenamefont
  {Besedin}, \citenamefont {Kuzovnikov}, \citenamefont {Mozaffari},
  \citenamefont {Balicas}, \citenamefont {Balakirev}, \citenamefont {Graf},
  \citenamefont {Prakapenka} \emph {et~al.}}]{drozdov2019superconductivity}%
  \BibitemOpen
  \bibfield  {author} {\bibinfo {author} {\bibfnamefont {A.}~\bibnamefont
  {Drozdov}}, \bibinfo {author} {\bibfnamefont {P.}~\bibnamefont {Kong}},
  \bibinfo {author} {\bibfnamefont {V.}~\bibnamefont {Minkov}}, \bibinfo
  {author} {\bibfnamefont {S.}~\bibnamefont {Besedin}}, \bibinfo {author}
  {\bibfnamefont {M.}~\bibnamefont {Kuzovnikov}}, \bibinfo {author}
  {\bibfnamefont {S.}~\bibnamefont {Mozaffari}}, \bibinfo {author}
  {\bibfnamefont {L.}~\bibnamefont {Balicas}}, \bibinfo {author} {\bibfnamefont
  {F.}~\bibnamefont {Balakirev}}, \bibinfo {author} {\bibfnamefont
  {D.}~\bibnamefont {Graf}}, \bibinfo {author} {\bibfnamefont {V.}~\bibnamefont
  {Prakapenka}}, \emph {et~al.},\ }\bibfield  {title} {\bibinfo {title}
  {{Superconductivity at 250 K in lanthanum hydride under high pressures}},\
  }\href {https://doi.org/10.1038/s41586-019-1201-8} {\bibfield  {journal}
  {\bibinfo  {journal} {Nature}\ }\textbf {\bibinfo {volume} {569}},\ \bibinfo
  {pages} {528} (\bibinfo {year} {2019})}\BibitemShut {NoStop}%
\bibitem [{\citenamefont {J{\'e}rome}(1991)}]{TMTCF}%
  \BibitemOpen
  \bibfield  {author} {\bibinfo {author} {\bibfnamefont {D.}~\bibnamefont
  {J{\'e}rome}},\ }\bibfield  {title} {\bibinfo {title} {{The physics of
  organic superconductors}},\ }\href
  {https://doi.org/10.1126/science.252.5012.1509} {\bibfield  {journal}
  {\bibinfo  {journal} {Science}\ }\textbf {\bibinfo {volume} {252}},\ \bibinfo
  {pages} {1509} (\bibinfo {year} {1991})}\BibitemShut {NoStop}%
\bibitem [{\citenamefont {Ishiguro}\ \emph {et~al.}(1998)\citenamefont
  {Ishiguro}, \citenamefont {Yamaji},\ and\ \citenamefont
  {Saito}}]{organic_superconductors}%
  \BibitemOpen
  \bibfield  {author} {\bibinfo {author} {\bibfnamefont {T.}~\bibnamefont
  {Ishiguro}}, \bibinfo {author} {\bibfnamefont {K.}~\bibnamefont {Yamaji}},\
  and\ \bibinfo {author} {\bibfnamefont {G.}~\bibnamefont {Saito}},\ }\href
  {https://doi.org/https://doi.org/10.1007/978-3-642-97190-7} {\emph {\bibinfo
  {title} {{Organic Superconductors}}}},\ \bibinfo {edition} {2nd}\ ed.\
  (\bibinfo  {publisher} {Springer},\ \bibinfo {address} {Berlin},\ \bibinfo
  {year} {1998})\BibitemShut {NoStop}%
\bibitem [{\citenamefont {Pashkin}\ \emph {et~al.}(2009)\citenamefont
  {Pashkin}, \citenamefont {Dressel}, \citenamefont {Ebbinghaus}, \citenamefont
  {Hanfland},\ and\ \citenamefont {Kuntscher}}]{TM_st_transition}%
  \BibitemOpen
  \bibfield  {author} {\bibinfo {author} {\bibfnamefont {A.}~\bibnamefont
  {Pashkin}}, \bibinfo {author} {\bibfnamefont {M.}~\bibnamefont {Dressel}},
  \bibinfo {author} {\bibfnamefont {S.~G.}\ \bibnamefont {Ebbinghaus}},
  \bibinfo {author} {\bibfnamefont {M.}~\bibnamefont {Hanfland}},\ and\
  \bibinfo {author} {\bibfnamefont {C.~A.}\ \bibnamefont {Kuntscher}},\
  }\bibfield  {title} {\bibinfo {title} {{Pressure-induced structural phase
  transition in the Bechgaard-Fabre salts}},\ }\href
  {https://doi.org/https://doi.org/10.1016/j.synthmet.2009.07.039} {\bibfield
  {journal} {\bibinfo  {journal} {Synth. Met.}\ }\textbf {\bibinfo {volume}
  {159}},\ \bibinfo {pages} {2097} (\bibinfo {year} {2009})}\BibitemShut
  {NoStop}%
\bibitem [{\citenamefont {Cui}\ \emph {et~al.}(2021)\citenamefont {Cui},
  \citenamefont {Yeung}, \citenamefont {Kawasugi}, \citenamefont {Minamidate},
  \citenamefont {Saunders},\ and\ \citenamefont {Kato}}]{organic_xray}%
  \BibitemOpen
  \bibfield  {author} {\bibinfo {author} {\bibfnamefont {H.}~\bibnamefont
  {Cui}}, \bibinfo {author} {\bibfnamefont {H.~H.-M.}\ \bibnamefont {Yeung}},
  \bibinfo {author} {\bibfnamefont {Y.}~\bibnamefont {Kawasugi}}, \bibinfo
  {author} {\bibfnamefont {T.}~\bibnamefont {Minamidate}}, \bibinfo {author}
  {\bibfnamefont {L.~K.}\ \bibnamefont {Saunders}},\ and\ \bibinfo {author}
  {\bibfnamefont {R.}~\bibnamefont {Kato}},\ }\bibfield  {title} {\bibinfo
  {title} {{High-pressure crystal structure and unusual magnetoresistance of a
  single-component molecular conductor [Pd (dddt)$_2$](dddt= 5, 6-dihydro-1,
  4-dithiin-2, 3-dithiolate)}},\ }\href
  {https://doi.org/https://doi.org/10.3390/cryst11050534} {\bibfield  {journal}
  {\bibinfo  {journal} {Crystals}\ }\textbf {\bibinfo {volume} {11}},\ \bibinfo
  {pages} {534} (\bibinfo {year} {2021})}\BibitemShut {NoStop}%
\bibitem [{\citenamefont {Iida}\ \emph {et~al.}(2021)\citenamefont {Iida},
  \citenamefont {Sawada}, \citenamefont {Sasaki}, \citenamefont {Tsuchiya},
  \citenamefont {Minamidate}, \citenamefont {Matsunaga}, \citenamefont
  {Kawamoto},\ and\ \citenamefont {Nomura}}]{AuBr2_AP_C-SDW}%
  \BibitemOpen
  \bibfield  {author} {\bibinfo {author} {\bibfnamefont {Y.}~\bibnamefont
  {Iida}}, \bibinfo {author} {\bibfnamefont {M.}~\bibnamefont {Sawada}},
  \bibinfo {author} {\bibfnamefont {Y.}~\bibnamefont {Sasaki}}, \bibinfo
  {author} {\bibfnamefont {T.}~\bibnamefont {Tsuchiya}}, \bibinfo {author}
  {\bibfnamefont {T.}~\bibnamefont {Minamidate}}, \bibinfo {author}
  {\bibfnamefont {N.}~\bibnamefont {Matsunaga}}, \bibinfo {author}
  {\bibfnamefont {A.}~\bibnamefont {Kawamoto}},\ and\ \bibinfo {author}
  {\bibfnamefont {K.}~\bibnamefont {Nomura}},\ }\bibfield  {title} {\bibinfo
  {title} {{Spin density wave in the strongly dimerized quasi-one-dimensional
  organic conductor (DMET-TTF)${}_{2}{\mathrm{AuBr}}_{2}$}},\ }\href
  {https://doi.org/10.1103/PhysRevB.104.184409} {\bibfield  {journal} {\bibinfo
   {journal} {Phys. Rev. B}\ }\textbf {\bibinfo {volume} {104}},\ \bibinfo
  {pages} {184409} (\bibinfo {year} {2021})}\BibitemShut {NoStop}%
\bibitem [{\citenamefont {Kanoda}\ and\ \citenamefont
  {Kato}(2011)}]{Kanoda_ARCMP2011}%
  \BibitemOpen
  \bibfield  {author} {\bibinfo {author} {\bibfnamefont {K.}~\bibnamefont
  {Kanoda}}\ and\ \bibinfo {author} {\bibfnamefont {R.}~\bibnamefont {Kato}},\
  }\bibfield  {title} {\bibinfo {title} {Mott physics in organic conductors
  with triangular lattices},\ }\href
  {https://doi.org/10.1146/annurev-conmatphys-062910-140521} {\bibfield
  {journal} {\bibinfo  {journal} {Annu. Rev. Condens. Matter Phys.}\ }\textbf
  {\bibinfo {volume} {2}},\ \bibinfo {pages} {167} (\bibinfo {year}
  {2011})}\BibitemShut {NoStop}%
\bibitem [{\citenamefont {Powell}\ and\ \citenamefont
  {McKenzie}(2006)}]{Powell_review2006}%
  \BibitemOpen
  \bibfield  {author} {\bibinfo {author} {\bibfnamefont {B.~J.}\ \bibnamefont
  {Powell}}\ and\ \bibinfo {author} {\bibfnamefont {R.~H.}\ \bibnamefont
  {McKenzie}},\ }\bibfield  {title} {\bibinfo {title} {{Strong electronic
  correlations in superconducting organic charge transfer salts}},\ }\href
  {https://doi.org/10.1088/0953-8984/18/45/r03} {\bibfield  {journal} {\bibinfo
   {journal} {J. Phys.: Cond. Matter.}\ }\textbf {\bibinfo {volume} {18}},\
  \bibinfo {pages} {R827} (\bibinfo {year} {2006})}\BibitemShut {NoStop}%
\bibitem [{\citenamefont {Miyagawa}\ \emph {et~al.}(1995)\citenamefont
  {Miyagawa}, \citenamefont {Kawamoto}, \citenamefont {Nakazawa},\ and\
  \citenamefont {Kanoda}}]{kET_AF}%
  \BibitemOpen
  \bibfield  {author} {\bibinfo {author} {\bibfnamefont {K.}~\bibnamefont
  {Miyagawa}}, \bibinfo {author} {\bibfnamefont {A.}~\bibnamefont {Kawamoto}},
  \bibinfo {author} {\bibfnamefont {Y.}~\bibnamefont {Nakazawa}},\ and\
  \bibinfo {author} {\bibfnamefont {K.}~\bibnamefont {Kanoda}},\ }\bibfield
  {title} {\bibinfo {title} {{Antiferromagnetic Ordering and Spin Structure in
  the Organic Conductor,$\kappa$-(BEDT-TTF)$_{2}$Cu[N(CN)$_{2}$]Cl}},\ }\href
  {https://doi.org/10.1103/PhysRevLett.75.1174} {\bibfield  {journal} {\bibinfo
   {journal} {Phys. Rev. Lett.}\ }\textbf {\bibinfo {volume} {75}},\ \bibinfo
  {pages} {1174} (\bibinfo {year} {1995})}\BibitemShut {NoStop}%
\bibitem [{\citenamefont {Yasin}\ \emph {et~al.}(2011)\citenamefont {Yasin},
  \citenamefont {Dumm}, \citenamefont {Salameh}, \citenamefont {Batail},
  \citenamefont {Me{\'z}i{\`e}re},\ and\ \citenamefont {Dressel}}]{k-Br}%
  \BibitemOpen
  \bibfield  {author} {\bibinfo {author} {\bibfnamefont {S.}~\bibnamefont
  {Yasin}}, \bibinfo {author} {\bibfnamefont {M.}~\bibnamefont {Dumm}},
  \bibinfo {author} {\bibfnamefont {B.}~\bibnamefont {Salameh}}, \bibinfo
  {author} {\bibfnamefont {P.}~\bibnamefont {Batail}}, \bibinfo {author}
  {\bibfnamefont {C.}~\bibnamefont {Me{\'z}i{\`e}re}},\ and\ \bibinfo {author}
  {\bibfnamefont {M.}~\bibnamefont {Dressel}},\ }\bibfield  {title} {\bibinfo
  {title} {{Transport studies at the Mott transition of the two-dimensional
  organic metal $\kappa$-(BEDT-TTF)$_2$Cu[N(CN)$_2$]Br$_x$Cl$_{1-x}$}},\ }\href
  {https://doi.org/https://doi.org/10.1140/epjb/e2010-10743-2} {\bibfield
  {journal} {\bibinfo  {journal} {The European Physical Journal B}\ }\textbf
  {\bibinfo {volume} {79}},\ \bibinfo {pages} {383} (\bibinfo {year}
  {2011})}\BibitemShut {NoStop}%
\bibitem [{\citenamefont {Hodges}\ \emph {et~al.}(2002)\citenamefont {Hodges},
  \citenamefont {Sidis}, \citenamefont {Bourges}, \citenamefont {Mirebeau},
  \citenamefont {Hennion},\ and\ \citenamefont {Chaud}}]{YBCO}%
  \BibitemOpen
  \bibfield  {author} {\bibinfo {author} {\bibfnamefont {J.~A.}\ \bibnamefont
  {Hodges}}, \bibinfo {author} {\bibfnamefont {Y.}~\bibnamefont {Sidis}},
  \bibinfo {author} {\bibfnamefont {P.}~\bibnamefont {Bourges}}, \bibinfo
  {author} {\bibfnamefont {I.}~\bibnamefont {Mirebeau}}, \bibinfo {author}
  {\bibfnamefont {M.}~\bibnamefont {Hennion}},\ and\ \bibinfo {author}
  {\bibfnamefont {X.}~\bibnamefont {Chaud}},\ }\bibfield  {title} {\bibinfo
  {title} {{Antiferromagnetic ordering in a 90 K copper oxide
  superconductor}},\ }\href {https://doi.org/10.1103/PhysRevB.66.020501}
  {\bibfield  {journal} {\bibinfo  {journal} {Phys. Rev. B}\ }\textbf {\bibinfo
  {volume} {66}},\ \bibinfo {pages} {020501} (\bibinfo {year}
  {2002})}\BibitemShut {NoStop}%
\bibitem [{\citenamefont {Kobayashi}\ \emph {et~al.}(1986)\citenamefont
  {Kobayashi}, \citenamefont {Kato}, \citenamefont {Kobayashi}, \citenamefont
  {Saito}, \citenamefont {Tokumoto}, \citenamefont {Anzai},\ and\ \citenamefont
  {Ishiguro}}]{ICl2_structure}%
  \BibitemOpen
  \bibfield  {author} {\bibinfo {author} {\bibfnamefont {H.}~\bibnamefont
  {Kobayashi}}, \bibinfo {author} {\bibfnamefont {R.}~\bibnamefont {Kato}},
  \bibinfo {author} {\bibfnamefont {A.}~\bibnamefont {Kobayashi}}, \bibinfo
  {author} {\bibfnamefont {G.}~\bibnamefont {Saito}}, \bibinfo {author}
  {\bibfnamefont {M.}~\bibnamefont {Tokumoto}}, \bibinfo {author}
  {\bibfnamefont {H.}~\bibnamefont {Anzai}},\ and\ \bibinfo {author}
  {\bibfnamefont {T.}~\bibnamefont {Ishiguro}},\ }\bibfield  {title} {\bibinfo
  {title} {{The crystal structure of $\beta^{\prime}$-(BEDT-TTF)$_2$ICl$_2$. A
  modification of the organic superconductor, $\beta$-(BEDT-TTF)$_2$I$_3$}},\
  }\href {https://doi.org/https://doi.org/10.1246/cl.1986.89} {\bibfield
  {journal} {\bibinfo  {journal} {Chemistry Letters}\ }\textbf {\bibinfo
  {volume} {15}},\ \bibinfo {pages} {89} (\bibinfo {year} {1986})}\BibitemShut
  {NoStop}%
\bibitem [{\citenamefont {Taniguchi}\ \emph {et~al.}(2003)\citenamefont
  {Taniguchi}, \citenamefont {Miyashita}, \citenamefont {Uchiyama},
  \citenamefont {Satoh}, \citenamefont {M{\^o}ri}, \citenamefont {Okamoto},
  \citenamefont {Miyagawa}, \citenamefont {Kanoda}, \citenamefont {Hedo},\ and\
  \citenamefont {Uwatoko}}]{ICl2_Taniguchi}%
  \BibitemOpen
  \bibfield  {author} {\bibinfo {author} {\bibfnamefont {H.}~\bibnamefont
  {Taniguchi}}, \bibinfo {author} {\bibfnamefont {M.}~\bibnamefont
  {Miyashita}}, \bibinfo {author} {\bibfnamefont {K.}~\bibnamefont {Uchiyama}},
  \bibinfo {author} {\bibfnamefont {K.}~\bibnamefont {Satoh}}, \bibinfo
  {author} {\bibfnamefont {N.}~\bibnamefont {M{\^o}ri}}, \bibinfo {author}
  {\bibfnamefont {H.}~\bibnamefont {Okamoto}}, \bibinfo {author} {\bibfnamefont
  {K.}~\bibnamefont {Miyagawa}}, \bibinfo {author} {\bibfnamefont
  {K.}~\bibnamefont {Kanoda}}, \bibinfo {author} {\bibfnamefont
  {M.}~\bibnamefont {Hedo}},\ and\ \bibinfo {author} {\bibfnamefont
  {Y.}~\bibnamefont {Uwatoko}},\ }\bibfield  {title} {\bibinfo {title}
  {{Superconductivity at 14.2 K in layered organics under extreme pressure}},\
  }\href {https://doi.org/https://doi.org/10.1143/JPSJ.72.468} {\bibfield
  {journal} {\bibinfo  {journal} {J.Phys. Soc. Jpn.}\ }\textbf {\bibinfo
  {volume} {72}},\ \bibinfo {pages} {468} (\bibinfo {year} {2003})}\BibitemShut
  {NoStop}%
\bibitem [{\citenamefont {Yoneyama}\ \emph {et~al.}(1997)\citenamefont
  {Yoneyama}, \citenamefont {Miyazaki}, \citenamefont {Enoki},\ and\
  \citenamefont {Saito}}]{ICl2_TN}%
  \BibitemOpen
  \bibfield  {author} {\bibinfo {author} {\bibfnamefont {N.}~\bibnamefont
  {Yoneyama}}, \bibinfo {author} {\bibfnamefont {A.}~\bibnamefont {Miyazaki}},
  \bibinfo {author} {\bibfnamefont {T.}~\bibnamefont {Enoki}},\ and\ \bibinfo
  {author} {\bibfnamefont {G.}~\bibnamefont {Saito}},\ }\bibfield  {title}
  {\bibinfo {title} {{Magnetic properties of (BEDT-TTF)$_2$X with localized
  spins}},\ }\href
  {https://doi.org/https://doi.org/10.1016/S0379-6779(97)81011-9} {\bibfield
  {journal} {\bibinfo  {journal} {Synth. Met.}\ }\textbf {\bibinfo {volume}
  {86}},\ \bibinfo {pages} {2029} (\bibinfo {year} {1997})}\BibitemShut
  {NoStop}%
\bibitem [{\citenamefont {Eto}\ and\ \citenamefont
  {Kawamoto}(2010)}]{Icl2_NMR}%
  \BibitemOpen
  \bibfield  {author} {\bibinfo {author} {\bibfnamefont {Y.}~\bibnamefont
  {Eto}}\ and\ \bibinfo {author} {\bibfnamefont {A.}~\bibnamefont {Kawamoto}},\
  }\bibfield  {title} {\bibinfo {title} {{Antiferromagnetic phase in
  ${\ensuremath{\beta}}^{\ensuremath{'}}\text{\ensuremath{-}}{(\text{BEDT-TTF})}_{2}{\text{ICl}}_{2}$
  under pressure as seen via $^{13}\text{C}$ NMR}},\ }\href
  {https://doi.org/10.1103/PhysRevB.81.020512} {\bibfield  {journal} {\bibinfo
  {journal} {Phys. Rev. B}\ }\textbf {\bibinfo {volume} {81}},\ \bibinfo
  {pages} {020512} (\bibinfo {year} {2010})}\BibitemShut {NoStop}%
\bibitem [{sup()}]{supplementary}%
  \BibitemOpen
  \href@noop {} {}\bibinfo {howpublished} {See Supplemental Material at
  \url{URL_will_be_inserted_by_publisher} for details of synthesis and x-ray
  diffraction experiments, resistivity measurements, $^{13}$C-NMR measurements,
  and derivation of the $ab$-$initio$ effective Hamiltonians, which includes
  Refs.~\cite{XT,XL,olex2,41,7373,CAC,PCAC,ICl2_Taniguchi,AuCl2_Taniguchi,coupling_1,coupling_2,AuBr2_AP_C-SDW,AuBr2_AP_C-SDW,kET_AF,ICl2_TN,AuBr2_AP_C-SDW,2DXY,QE,Hamann_ONCV2013,
  Schlipf_CPC2015,GGA_PBE,PhysRevB.70.195104, Imada_JPSJ2010,Shinaoka2012,
  PhysRevResearch.2.032072, PhysRevResearch.3.043224, Ido2022,
  PhysRevB.105.205123, PhysRevB.102.235116, yoshimi2022,RESPACK, FC70_FC77,
  Fluorinert, Daphne7373}}\BibitemShut {NoStop}%
\bibitem [{\citenamefont {Mori}(1998)}]{RARB}%
  \BibitemOpen
  \bibfield  {author} {\bibinfo {author} {\bibfnamefont {T.}~\bibnamefont
  {Mori}},\ }\bibfield  {title} {\bibinfo {title} {{Structural genealogy of
  BEDT-TTF-based organic conductors I. Parallel molecules: $\beta$ and
  $\beta^{\prime\prime}$ phases}},\ }\href
  {https://doi.org/https://doi.org/10.1246/bcsj.71.2509} {\bibfield  {journal}
  {\bibinfo  {journal} {Bull. Chem. Soc. J.}\ }\textbf {\bibinfo {volume}
  {71}},\ \bibinfo {pages} {2509} (\bibinfo {year} {1998})}\BibitemShut
  {NoStop}%
\bibitem [{\citenamefont {Svenstrup}\ \emph {et~al.}(1994)\citenamefont
  {Svenstrup}, \citenamefont {Rasmussen}, \citenamefont {Hansen},\ and\
  \citenamefont {Becher}}]{coupling_1}%
  \BibitemOpen
  \bibfield  {author} {\bibinfo {author} {\bibfnamefont {N.}~\bibnamefont
  {Svenstrup}}, \bibinfo {author} {\bibfnamefont {K.~M.}\ \bibnamefont
  {Rasmussen}}, \bibinfo {author} {\bibfnamefont {T.~K.}\ \bibnamefont
  {Hansen}},\ and\ \bibinfo {author} {\bibfnamefont {J.}~\bibnamefont
  {Becher}},\ }\bibfield  {title} {\bibinfo {title} {{The chemistry of TTFTT;
  1: new efficient synthesis and reactions of tetrathiafulvalene-2, 3, 6,
  7-tetrathiolate (TTFTT): an important building block in TTF-syntheses}},\
  }\href {https://doi.org/10.1055/s-1994-25580} {\bibfield  {journal} {\bibinfo
   {journal} {Synthesis}\ }\textbf {\bibinfo {volume} {1994}},\ \bibinfo
  {pages} {809} (\bibinfo {year} {1994})}\BibitemShut {NoStop}%
\bibitem [{\citenamefont {Hirose}\ \emph {et~al.}(2012)\citenamefont {Hirose},
  \citenamefont {Misawa},\ and\ \citenamefont {Kawamoto}}]{coupling_2}%
  \BibitemOpen
  \bibfield  {author} {\bibinfo {author} {\bibfnamefont {S.}~\bibnamefont
  {Hirose}}, \bibinfo {author} {\bibfnamefont {M.}~\bibnamefont {Misawa}},\
  and\ \bibinfo {author} {\bibfnamefont {A.}~\bibnamefont {Kawamoto}},\
  }\bibfield  {title} {\bibinfo {title} {{Magnetic and Electric Properties of
  Organic Conductors Probed by $^{13}$C-NMR Using Selective-Site Substituted
  Molecules}},\ }\href {https://doi.org/https://doi.org/10.3390/cryst2031034}
  {\bibfield  {journal} {\bibinfo  {journal} {Crystals}\ }\textbf {\bibinfo
  {volume} {2}},\ \bibinfo {pages} {1034} (\bibinfo {year} {2012})}\BibitemShut
  {NoStop}%
\bibitem [{\citenamefont {Cheng}\ \emph {et~al.}(2014)\citenamefont {Cheng},
  \citenamefont {Matsubayashi}, \citenamefont {Nagasaki}, \citenamefont
  {Hisada}, \citenamefont {Hirayama}, \citenamefont {Hedo}, \citenamefont
  {Kagi},\ and\ \citenamefont {Uwatoko}}]{PCAC}%
  \BibitemOpen
  \bibfield  {author} {\bibinfo {author} {\bibfnamefont {J.-G.}\ \bibnamefont
  {Cheng}}, \bibinfo {author} {\bibfnamefont {K.}~\bibnamefont {Matsubayashi}},
  \bibinfo {author} {\bibfnamefont {S.}~\bibnamefont {Nagasaki}}, \bibinfo
  {author} {\bibfnamefont {A.}~\bibnamefont {Hisada}}, \bibinfo {author}
  {\bibfnamefont {T.}~\bibnamefont {Hirayama}}, \bibinfo {author}
  {\bibfnamefont {M.}~\bibnamefont {Hedo}}, \bibinfo {author} {\bibfnamefont
  {H.}~\bibnamefont {Kagi}},\ and\ \bibinfo {author} {\bibfnamefont
  {Y.}~\bibnamefont {Uwatoko}},\ }\bibfield  {title} {\bibinfo {title}
  {{Integrated-fin gasket for palm cubic-anvil high pressure apparatus}},\
  }\href {https://doi.org/https://doi.org/10.1063/1.4896473} {\bibfield
  {journal} {\bibinfo  {journal} {Rev. Sci.Instrum.}\ }\textbf {\bibinfo
  {volume} {85}},\ \bibinfo {pages} {093907} (\bibinfo {year}
  {2014})}\BibitemShut {NoStop}%
\bibitem [{\citenamefont {Mori}\ \emph {et~al.}(2004)\citenamefont {Mori},
  \citenamefont {Takahashi},\ and\ \citenamefont {Takeshita}}]{CAC}%
  \BibitemOpen
  \bibfield  {author} {\bibinfo {author} {\bibfnamefont {N.}~\bibnamefont
  {Mori}}, \bibinfo {author} {\bibfnamefont {H.}~\bibnamefont {Takahashi}},\
  and\ \bibinfo {author} {\bibfnamefont {N.}~\bibnamefont {Takeshita}},\
  }\bibfield  {title} {\bibinfo {title} {{Low-temperature and high-pressure
  apparatus developed at ISSP, University of Tokyo}},\ }\href
  {https://doi.org/https://doi.org/10.1080/08957950410001661909} {\bibfield
  {journal} {\bibinfo  {journal} {High Pressure Research}\ }\textbf {\bibinfo
  {volume} {24}},\ \bibinfo {pages} {225} (\bibinfo {year} {2004})}\BibitemShut
  {NoStop}%
\bibitem [{\citenamefont {Kagawa}\ \emph {et~al.}(2005)\citenamefont {Kagawa},
  \citenamefont {Miyagawa},\ and\ \citenamefont {Kanoda}}]{Kagawa_Nature2005}%
  \BibitemOpen
  \bibfield  {author} {\bibinfo {author} {\bibfnamefont {F.}~\bibnamefont
  {Kagawa}}, \bibinfo {author} {\bibfnamefont {K.}~\bibnamefont {Miyagawa}},\
  and\ \bibinfo {author} {\bibfnamefont {K.}~\bibnamefont {Kanoda}},\
  }\bibfield  {title} {\bibinfo {title} {Unconventional critical behaviour in a
  quasi-two-dimensional organic conductor},\ }\href
  {https://doi.org/https://doi.org/10.1038/nature03806} {\bibfield  {journal}
  {\bibinfo  {journal} {Nature}\ }\textbf {\bibinfo {volume} {436}},\ \bibinfo
  {pages} {534} (\bibinfo {year} {2005})}\BibitemShut {NoStop}%
\bibitem [{\citenamefont {Furukawa}\ \emph {et~al.}(2015)\citenamefont
  {Furukawa}, \citenamefont {Miyagawa}, \citenamefont {Taniguchi},
  \citenamefont {Kato},\ and\ \citenamefont {Kanoda}}]{QCP1}%
  \BibitemOpen
  \bibfield  {author} {\bibinfo {author} {\bibfnamefont {T.}~\bibnamefont
  {Furukawa}}, \bibinfo {author} {\bibfnamefont {K.}~\bibnamefont {Miyagawa}},
  \bibinfo {author} {\bibfnamefont {H.}~\bibnamefont {Taniguchi}}, \bibinfo
  {author} {\bibfnamefont {R.}~\bibnamefont {Kato}},\ and\ \bibinfo {author}
  {\bibfnamefont {K.}~\bibnamefont {Kanoda}},\ }\bibfield  {title} {\bibinfo
  {title} {{Quantum criticality of Mott transition in organic materials}},\
  }\href {https://doi.org/https://doi.org/10.1038/nphys3235} {\bibfield
  {journal} {\bibinfo  {journal} {Nature Physics}\ }\textbf {\bibinfo {volume}
  {11}},\ \bibinfo {pages} {221} (\bibinfo {year} {2015})}\BibitemShut
  {NoStop}%
\bibitem [{\citenamefont {Furukawa}\ \emph {et~al.}(2018)\citenamefont
  {Furukawa}, \citenamefont {Kobashi}, \citenamefont {Kurosaki}, \citenamefont
  {Miyagawa},\ and\ \citenamefont {Kanoda}}]{QCP2}%
  \BibitemOpen
  \bibfield  {author} {\bibinfo {author} {\bibfnamefont {T.}~\bibnamefont
  {Furukawa}}, \bibinfo {author} {\bibfnamefont {K.}~\bibnamefont {Kobashi}},
  \bibinfo {author} {\bibfnamefont {Y.}~\bibnamefont {Kurosaki}}, \bibinfo
  {author} {\bibfnamefont {K.}~\bibnamefont {Miyagawa}},\ and\ \bibinfo
  {author} {\bibfnamefont {K.}~\bibnamefont {Kanoda}},\ }\bibfield  {title}
  {\bibinfo {title} {{Quasi-continuous transition from a Fermi liquid to a spin
  liquid in $\kappa$-(ET)$_2$Cu$_2$(CN)$_3$}},\ }\href
  {https://doi.org/https://doi.org/10.1038/s41467-017-02679-7} {\bibfield
  {journal} {\bibinfo  {journal} {Nature communications}\ }\textbf {\bibinfo
  {volume} {9}},\ \bibinfo {pages} {307} (\bibinfo {year} {2018})}\BibitemShut
  {NoStop}%
\bibitem [{\citenamefont {Aryasetiawan}\ \emph {et~al.}(2004)\citenamefont
  {Aryasetiawan}, \citenamefont {Imada}, \citenamefont {Georges}, \citenamefont
  {Kotliar}, \citenamefont {Biermann},\ and\ \citenamefont
  {Lichtenstein}}]{PhysRevB.70.195104}%
  \BibitemOpen
  \bibfield  {author} {\bibinfo {author} {\bibfnamefont {F.}~\bibnamefont
  {Aryasetiawan}}, \bibinfo {author} {\bibfnamefont {M.}~\bibnamefont {Imada}},
  \bibinfo {author} {\bibfnamefont {A.}~\bibnamefont {Georges}}, \bibinfo
  {author} {\bibfnamefont {G.}~\bibnamefont {Kotliar}}, \bibinfo {author}
  {\bibfnamefont {S.}~\bibnamefont {Biermann}},\ and\ \bibinfo {author}
  {\bibfnamefont {A.~I.}\ \bibnamefont {Lichtenstein}},\ }\bibfield  {title}
  {\bibinfo {title} {Frequency-dependent local interactions and low-energy
  effective models from electronic structure calculations},\ }\href
  {https://doi.org/10.1103/PhysRevB.70.195104} {\bibfield  {journal} {\bibinfo
  {journal} {Phys. Rev. B}\ }\textbf {\bibinfo {volume} {70}},\ \bibinfo
  {pages} {195104} (\bibinfo {year} {2004})}\BibitemShut {NoStop}%
\bibitem [{\citenamefont {Imada}\ and\ \citenamefont
  {Miyake}(2010)}]{Imada_JPSJ2010}%
  \BibitemOpen
  \bibfield  {author} {\bibinfo {author} {\bibfnamefont {M.}~\bibnamefont
  {Imada}}\ and\ \bibinfo {author} {\bibfnamefont {T.}~\bibnamefont {Miyake}},\
  }\bibfield  {title} {\bibinfo {title} {Electronic structure calculation by
  first principles for strongly correlated electron systems},\ }\href
  {https://doi.org/10.1143/JPSJ.79.112001} {\bibfield  {journal} {\bibinfo
  {journal} {J. Phys. Soc. Jpn.}\ }\textbf {\bibinfo {volume} {79}},\ \bibinfo
  {pages} {112001} (\bibinfo {year} {2010})}\BibitemShut {NoStop}%
\bibitem [{\citenamefont {Nakamura}\ \emph {et~al.}(2021)\citenamefont
  {Nakamura}, \citenamefont {Yoshimoto}, \citenamefont {Nomura}, \citenamefont
  {Tadano}, \citenamefont {Kawamura}, \citenamefont {Kosugi}, \citenamefont
  {Yoshimi}, \citenamefont {Misawa},\ and\ \citenamefont {Motoyama}}]{RESPACK}%
  \BibitemOpen
  \bibfield  {author} {\bibinfo {author} {\bibfnamefont {K.}~\bibnamefont
  {Nakamura}}, \bibinfo {author} {\bibfnamefont {Y.}~\bibnamefont {Yoshimoto}},
  \bibinfo {author} {\bibfnamefont {Y.}~\bibnamefont {Nomura}}, \bibinfo
  {author} {\bibfnamefont {T.}~\bibnamefont {Tadano}}, \bibinfo {author}
  {\bibfnamefont {M.}~\bibnamefont {Kawamura}}, \bibinfo {author}
  {\bibfnamefont {T.}~\bibnamefont {Kosugi}}, \bibinfo {author} {\bibfnamefont
  {K.}~\bibnamefont {Yoshimi}}, \bibinfo {author} {\bibfnamefont
  {T.}~\bibnamefont {Misawa}},\ and\ \bibinfo {author} {\bibfnamefont
  {Y.}~\bibnamefont {Motoyama}},\ }\bibfield  {title} {\bibinfo {title}
  {{RESPACK: An ab initio tool for derivation of effective low-energy model of
  material}},\ }\href {https://doi.org/doi.org/10.1016/j.cpc.2020.107781}
  {\bibfield  {journal} {\bibinfo  {journal} {Comp. Phys. Comm.}\ }\textbf
  {\bibinfo {volume} {261}},\ \bibinfo {pages} {107781} (\bibinfo {year}
  {2021})}\BibitemShut {NoStop}%
\bibitem [{\citenamefont {Sheldrick}(2015{\natexlab{a}})}]{XT}%
  \BibitemOpen
  \bibfield  {author} {\bibinfo {author} {\bibfnamefont {G.~M.}\ \bibnamefont
  {Sheldrick}},\ }\bibfield  {title} {\bibinfo {title} {{Crystal structure
  refinement with SHELXL}},\ }\href
  {https://doi.org/https://doi.org/10.1107/S2053229614024218} {\bibfield
  {journal} {\bibinfo  {journal} {Acta crystallogr. C: Struct. Chem.}\ }\textbf
  {\bibinfo {volume} {71}},\ \bibinfo {pages} {3} (\bibinfo {year}
  {2015}{\natexlab{a}})}\BibitemShut {NoStop}%
\bibitem [{\citenamefont {Dolomanov}\ \emph {et~al.}(2009)\citenamefont
  {Dolomanov}, \citenamefont {Bourhis}, \citenamefont {Gildea}, \citenamefont
  {Howard},\ and\ \citenamefont {Puschmann}}]{olex2}%
  \BibitemOpen
  \bibfield  {author} {\bibinfo {author} {\bibfnamefont {O.~V.}\ \bibnamefont
  {Dolomanov}}, \bibinfo {author} {\bibfnamefont {L.~J.}\ \bibnamefont
  {Bourhis}}, \bibinfo {author} {\bibfnamefont {R.~J.}\ \bibnamefont {Gildea}},
  \bibinfo {author} {\bibfnamefont {J.~A.}\ \bibnamefont {Howard}},\ and\
  \bibinfo {author} {\bibfnamefont {H.}~\bibnamefont {Puschmann}},\ }\bibfield
  {title} {\bibinfo {title} {{OLEX2: a complete structure solution, refinement
  and analysis program}},\ }\href
  {https://doi.org/https://doi.org/10.1107/S0021889808042726} {\bibfield
  {journal} {\bibinfo  {journal} {J. Appl. Crystallogr.}\ }\textbf {\bibinfo
  {volume} {42}},\ \bibinfo {pages} {339} (\bibinfo {year} {2009})}\BibitemShut
  {NoStop}%
\bibitem [{\citenamefont {Sheldrick}(2015{\natexlab{b}})}]{XL}%
  \BibitemOpen
  \bibfield  {author} {\bibinfo {author} {\bibfnamefont {G.~M.}\ \bibnamefont
  {Sheldrick}},\ }\bibfield  {title} {\bibinfo {title} {Shelxt--integrated
  space-group and crystal-structure determination},\ }\href@noop {} {\bibfield
  {journal} {\bibinfo  {journal} {Acta Crystallographica Section A: Foundations
  and Advances}\ }\textbf {\bibinfo {volume} {71}},\ \bibinfo {pages} {3}
  (\bibinfo {year} {2015}{\natexlab{b}})}\BibitemShut {NoStop}%
\bibitem [{\citenamefont {Klotz}\ \emph {et~al.}(2009)\citenamefont {Klotz},
  \citenamefont {Paumier}, \citenamefont {Le~March},\ and\ \citenamefont
  {Munsch}}]{41}%
  \BibitemOpen
  \bibfield  {author} {\bibinfo {author} {\bibfnamefont {S.}~\bibnamefont
  {Klotz}}, \bibinfo {author} {\bibfnamefont {L.}~\bibnamefont {Paumier}},
  \bibinfo {author} {\bibfnamefont {G.}~\bibnamefont {Le~March}},\ and\
  \bibinfo {author} {\bibfnamefont {P.}~\bibnamefont {Munsch}},\ }\bibfield
  {title} {\bibinfo {title} {{The effect of temperature on the hydrostatic
  limit of 4: 1 methanol--ethanol under pressure}},\ }\href
  {https://doi.org/https://doi.org/10.1080/08957950903418194} {\bibfield
  {journal} {\bibinfo  {journal} {High Pressure Research}\ }\textbf {\bibinfo
  {volume} {29}},\ \bibinfo {pages} {649} (\bibinfo {year} {2009})}\BibitemShut
  {NoStop}%
\bibitem [{\citenamefont {Murata}\ \emph {et~al.}(1997)\citenamefont {Murata},
  \citenamefont {Yoshino}, \citenamefont {Yadav}, \citenamefont {Honda},\ and\
  \citenamefont {Shirakawa}}]{7373}%
  \BibitemOpen
  \bibfield  {author} {\bibinfo {author} {\bibfnamefont {K.}~\bibnamefont
  {Murata}}, \bibinfo {author} {\bibfnamefont {H.}~\bibnamefont {Yoshino}},
  \bibinfo {author} {\bibfnamefont {H.~O.}\ \bibnamefont {Yadav}}, \bibinfo
  {author} {\bibfnamefont {Y.}~\bibnamefont {Honda}},\ and\ \bibinfo {author}
  {\bibfnamefont {N.}~\bibnamefont {Shirakawa}},\ }\bibfield  {title} {\bibinfo
  {title} {{Pt resistor thermometry and pressure calibration in a clamped
  pressure cell with the medium, Daphne 7373}},\ }\href
  {https://doi.org/https://doi.org/10.1063/1.1148145} {\bibfield  {journal}
  {\bibinfo  {journal} {Rev. Sci.Instrum.}\ }\textbf {\bibinfo {volume} {68}},\
  \bibinfo {pages} {2490} (\bibinfo {year} {1997})}\BibitemShut {NoStop}%
\bibitem [{\citenamefont {Taniguchi}\ \emph {et~al.}(2005)\citenamefont
  {Taniguchi}, \citenamefont {Miyashita}, \citenamefont {Uchiyama},
  \citenamefont {Sato}, \citenamefont {Ishii}, \citenamefont {Satoh},
  \citenamefont {M{\^o}ri}, \citenamefont {Hedo},\ and\ \citenamefont
  {Uwatoko}}]{AuCl2_Taniguchi}%
  \BibitemOpen
  \bibfield  {author} {\bibinfo {author} {\bibfnamefont {H.}~\bibnamefont
  {Taniguchi}}, \bibinfo {author} {\bibfnamefont {M.}~\bibnamefont
  {Miyashita}}, \bibinfo {author} {\bibfnamefont {K.}~\bibnamefont {Uchiyama}},
  \bibinfo {author} {\bibfnamefont {R.}~\bibnamefont {Sato}}, \bibinfo {author}
  {\bibfnamefont {Y.}~\bibnamefont {Ishii}}, \bibinfo {author} {\bibfnamefont
  {K.}~\bibnamefont {Satoh}}, \bibinfo {author} {\bibfnamefont
  {N.}~\bibnamefont {M{\^o}ri}}, \bibinfo {author} {\bibfnamefont
  {M.}~\bibnamefont {Hedo}},\ and\ \bibinfo {author} {\bibfnamefont
  {Y.}~\bibnamefont {Uwatoko}},\ }\bibfield  {title} {\bibinfo {title}
  {{High-pressure study up to 9.9 GPa of organic Mott insulator,
  $\beta^{\prime}$-(BEDT-TTF)$_2$AuCl$_2$}},\ }\href
  {https://doi.org/https://doi.org/10.1143/JPSJ.74.1370} {\bibfield  {journal}
  {\bibinfo  {journal} {J.Phys. Soc. Jpn.}\ }\textbf {\bibinfo {volume} {74}},\
  \bibinfo {pages} {1370} (\bibinfo {year} {2005})}\BibitemShut {NoStop}%
\bibitem [{\citenamefont {Sidorov}\ and\ \citenamefont
  {Sadykov}(2005)}]{Fluorinert}%
  \BibitemOpen
  \bibfield  {author} {\bibinfo {author} {\bibfnamefont {V.}~\bibnamefont
  {Sidorov}}\ and\ \bibinfo {author} {\bibfnamefont {R.}~\bibnamefont
  {Sadykov}},\ }\bibfield  {title} {\bibinfo {title} {{Hydrostatic limits of
  Fluorinert liquids used for neutron and transport studies at high
  pressure}},\ }\href {https://doi.org/10.1088/0953-8984/17/40/002} {\bibfield
  {journal} {\bibinfo  {journal} {J. of Phys.: Cond. Matt}\ }\textbf {\bibinfo
  {volume} {17}},\ \bibinfo {pages} {S3005} (\bibinfo {year}
  {2005})}\BibitemShut {NoStop}%
\bibitem [{\citenamefont {Yokogawa}\ \emph {et~al.}(2007)\citenamefont
  {Yokogawa}, \citenamefont {Murata}, \citenamefont {Yoshino},\ and\
  \citenamefont {Aoyama}}]{Daphne7373}%
  \BibitemOpen
  \bibfield  {author} {\bibinfo {author} {\bibfnamefont {K.}~\bibnamefont
  {Yokogawa}}, \bibinfo {author} {\bibfnamefont {K.}~\bibnamefont {Murata}},
  \bibinfo {author} {\bibfnamefont {H.}~\bibnamefont {Yoshino}},\ and\ \bibinfo
  {author} {\bibfnamefont {S.}~\bibnamefont {Aoyama}},\ }\bibfield  {title}
  {\bibinfo {title} {{Solidification of high-pressure medium Daphne 7373}},\
  }\href {https://doi.org/10.1143/JJAP.46.3636} {\bibfield  {journal} {\bibinfo
   {journal} {J. J. Appl. Phys.}\ }\textbf {\bibinfo {volume} {46}},\ \bibinfo
  {pages} {3636} (\bibinfo {year} {2007})}\BibitemShut {NoStop}%
\bibitem [{\citenamefont {Bramwell}\ and\ \citenamefont
  {Holdsworth}(1993)}]{2DXY}%
  \BibitemOpen
  \bibfield  {author} {\bibinfo {author} {\bibfnamefont {S.}~\bibnamefont
  {Bramwell}}\ and\ \bibinfo {author} {\bibfnamefont {P.}~\bibnamefont
  {Holdsworth}},\ }\bibfield  {title} {\bibinfo {title} {{Magnetization and
  universal sub-critical behaviour in two-dimensional XY magnets}},\ }\href
  {https://doi.org/10.1088/0953-8984/5/4/004} {\bibfield  {journal} {\bibinfo
  {journal} {J. of Phys.: Cond. Matt.}\ }\textbf {\bibinfo {volume} {5}},\
  \bibinfo {pages} {L53} (\bibinfo {year} {1993})}\BibitemShut {NoStop}%
\bibitem [{\citenamefont {Giannozzi}\ \emph {et~al.}(2017)\citenamefont
  {Giannozzi}, \citenamefont {Andreussi}, \citenamefont {Brumme}, \citenamefont
  {Bunau}, \citenamefont {Nardelli}, \citenamefont {Calandra}, \citenamefont
  {Car}, \citenamefont {Cavazzoni}, \citenamefont {Ceresoli}, \citenamefont
  {Cococcioni} \emph {et~al.}}]{QE}%
  \BibitemOpen
  \bibfield  {author} {\bibinfo {author} {\bibfnamefont {P.}~\bibnamefont
  {Giannozzi}}, \bibinfo {author} {\bibfnamefont {O.}~\bibnamefont
  {Andreussi}}, \bibinfo {author} {\bibfnamefont {T.}~\bibnamefont {Brumme}},
  \bibinfo {author} {\bibfnamefont {O.}~\bibnamefont {Bunau}}, \bibinfo
  {author} {\bibfnamefont {M.~B.}\ \bibnamefont {Nardelli}}, \bibinfo {author}
  {\bibfnamefont {M.}~\bibnamefont {Calandra}}, \bibinfo {author}
  {\bibfnamefont {R.}~\bibnamefont {Car}}, \bibinfo {author} {\bibfnamefont
  {C.}~\bibnamefont {Cavazzoni}}, \bibinfo {author} {\bibfnamefont
  {D.}~\bibnamefont {Ceresoli}}, \bibinfo {author} {\bibfnamefont
  {M.}~\bibnamefont {Cococcioni}}, \emph {et~al.},\ }\bibfield  {title}
  {\bibinfo {title} {{Advanced capabilities for materials modelling with
  Quantum ESPRESSO}},\ }\href {https://doi.org/10.1088/1361-648X/aa8f79}
  {\bibfield  {journal} {\bibinfo  {journal} {J. of Phys.: Cond. Matt.}\
  }\textbf {\bibinfo {volume} {29}},\ \bibinfo {pages} {465901} (\bibinfo
  {year} {2017})}\BibitemShut {NoStop}%
\bibitem [{\citenamefont {Hamann}(2013)}]{Hamann_ONCV2013}%
  \BibitemOpen
  \bibfield  {author} {\bibinfo {author} {\bibfnamefont {D.~R.}\ \bibnamefont
  {Hamann}},\ }\bibfield  {title} {\bibinfo {title} {{Optimized norm-conserving
  Vanderbilt pseudopotentials}},\ }\href
  {https://doi.org/10.1103/PhysRevB.88.085117} {\bibfield  {journal} {\bibinfo
  {journal} {Phys. Rev. B}\ }\textbf {\bibinfo {volume} {88}},\ \bibinfo
  {pages} {085117} (\bibinfo {year} {2013})}\BibitemShut {NoStop}%
\bibitem [{\citenamefont {Schlipf}\ and\ \citenamefont
  {Gygi}(2015)}]{Schlipf_CPC2015}%
  \BibitemOpen
  \bibfield  {author} {\bibinfo {author} {\bibfnamefont {M.}~\bibnamefont
  {Schlipf}}\ and\ \bibinfo {author} {\bibfnamefont {F.}~\bibnamefont {Gygi}},\
  }\bibfield  {title} {\bibinfo {title} {{Optimization algorithm for the
  generation of ONCV pseudopotentials}},\ }\href
  {https://doi.org/https://doi.org/10.1016/j.cpc.2015.05.011} {\bibfield
  {journal} {\bibinfo  {journal} {Comput. Phys. Commun.}\ }\textbf {\bibinfo
  {volume} {196}},\ \bibinfo {pages} {36} (\bibinfo {year} {2015})}\BibitemShut
  {NoStop}%
\bibitem [{\citenamefont {Perdew}\ \emph {et~al.}(1996)\citenamefont {Perdew},
  \citenamefont {Burke},\ and\ \citenamefont {Ernzerhof}}]{GGA_PBE}%
  \BibitemOpen
  \bibfield  {author} {\bibinfo {author} {\bibfnamefont {J.~P.}\ \bibnamefont
  {Perdew}}, \bibinfo {author} {\bibfnamefont {K.}~\bibnamefont {Burke}},\ and\
  \bibinfo {author} {\bibfnamefont {M.}~\bibnamefont {Ernzerhof}},\ }\bibfield
  {title} {\bibinfo {title} {Generalized gradient approximation made simple},\
  }\href {https://doi.org/10.1103/PhysRevLett.77.3865} {\bibfield  {journal}
  {\bibinfo  {journal} {Phys. Rev. Lett.}\ }\textbf {\bibinfo {volume} {77}},\
  \bibinfo {pages} {3865} (\bibinfo {year} {1996})}\BibitemShut {NoStop}%
\bibitem [{\citenamefont {Shinaoka}\ \emph {et~al.}(2012)\citenamefont
  {Shinaoka}, \citenamefont {Misawa}, \citenamefont {Nakamura},\ and\
  \citenamefont {Imada}}]{Shinaoka2012}%
  \BibitemOpen
  \bibfield  {author} {\bibinfo {author} {\bibfnamefont {H.}~\bibnamefont
  {Shinaoka}}, \bibinfo {author} {\bibfnamefont {T.}~\bibnamefont {Misawa}},
  \bibinfo {author} {\bibfnamefont {K.}~\bibnamefont {Nakamura}},\ and\
  \bibinfo {author} {\bibfnamefont {M.}~\bibnamefont {Imada}},\ }\bibfield
  {title} {\bibinfo {title} {{Mott Transition and Phase Diagram of
  $\kappa$-(BEDT-TTF)$_2$Cu(NCS)$_2$ Studied by Two-Dimensional Model Derived
  from Ab initio Method}},\ }\href {https://doi.org/10.1143/JPSJ.81.034701}
  {\bibfield  {journal} {\bibinfo  {journal} {J. Phys. Soc. Jpn}\ }\textbf
  {\bibinfo {volume} {81}},\ \bibinfo {pages} {034701} (\bibinfo {year}
  {2012})}\BibitemShut {NoStop}%
\bibitem [{\citenamefont {Misawa}\ \emph {et~al.}(2020)\citenamefont {Misawa},
  \citenamefont {Yoshimi},\ and\ \citenamefont
  {Tsumuraya}}]{PhysRevResearch.2.032072}%
  \BibitemOpen
  \bibfield  {author} {\bibinfo {author} {\bibfnamefont {T.}~\bibnamefont
  {Misawa}}, \bibinfo {author} {\bibfnamefont {K.}~\bibnamefont {Yoshimi}},\
  and\ \bibinfo {author} {\bibfnamefont {T.}~\bibnamefont {Tsumuraya}},\
  }\bibfield  {title} {\bibinfo {title} {{Electronic correlation and
  geometrical frustration in molecular solids: A systematic ab initio study of
  ${\ensuremath{\beta}}^{\ensuremath{'}}\text{\ensuremath{-}}X{[\mathrm{Pd}{(\mathrm{dmit})}_{2}]}_{2}$}},\
  }\href {https://doi.org/10.1103/PhysRevResearch.2.032072} {\bibfield
  {journal} {\bibinfo  {journal} {Phys. Rev. Research}\ }\textbf {\bibinfo
  {volume} {2}},\ \bibinfo {pages} {032072} (\bibinfo {year}
  {2020})}\BibitemShut {NoStop}%
\bibitem [{\citenamefont {Yoshimi}\ \emph {et~al.}(2021)\citenamefont
  {Yoshimi}, \citenamefont {Tsumuraya},\ and\ \citenamefont
  {Misawa}}]{PhysRevResearch.3.043224}%
  \BibitemOpen
  \bibfield  {author} {\bibinfo {author} {\bibfnamefont {K.}~\bibnamefont
  {Yoshimi}}, \bibinfo {author} {\bibfnamefont {T.}~\bibnamefont {Tsumuraya}},\
  and\ \bibinfo {author} {\bibfnamefont {T.}~\bibnamefont {Misawa}},\
  }\bibfield  {title} {\bibinfo {title} {Ab initio derivation and exact
  diagonalization analysis of low-energy effective hamiltonians for
  ${\ensuremath{\beta}}^{\ensuremath{'}}\text{\ensuremath{-}}\mathrm{X}{[\mathrm{Pd}{(\mathrm{dmit})}_{2}]}_{2}$},\
  }\href {https://doi.org/10.1103/PhysRevResearch.3.043224} {\bibfield
  {journal} {\bibinfo  {journal} {Phys. Rev. Research}\ }\textbf {\bibinfo
  {volume} {3}},\ \bibinfo {pages} {043224} (\bibinfo {year}
  {2021})}\BibitemShut {NoStop}%
\bibitem [{\citenamefont {Ido}\ \emph {et~al.}(2022)\citenamefont {Ido},
  \citenamefont {Yoshimi}, \citenamefont {Misawa},\ and\ \citenamefont
  {Imada}}]{Ido2022}%
  \BibitemOpen
  \bibfield  {author} {\bibinfo {author} {\bibfnamefont {K.}~\bibnamefont
  {Ido}}, \bibinfo {author} {\bibfnamefont {K.}~\bibnamefont {Yoshimi}},
  \bibinfo {author} {\bibfnamefont {T.}~\bibnamefont {Misawa}},\ and\ \bibinfo
  {author} {\bibfnamefont {M.}~\bibnamefont {Imada}},\ }\bibfield  {title}
  {\bibinfo {title} {Unconventional dual 1d--2d quantum spin liquid revealed by
  ab initio studies on organic solids family},\ }\href
  {https://doi.org/10.1038/s41535-022-00452-8} {\bibfield  {journal} {\bibinfo
  {journal} {npj Quantum Mater.}\ }\textbf {\bibinfo {volume} {7}},\ \bibinfo
  {pages} {48} (\bibinfo {year} {2022})}\BibitemShut {NoStop}%
\bibitem [{\citenamefont {Ohki}\ \emph {et~al.}(2022)\citenamefont {Ohki},
  \citenamefont {Yoshimi},\ and\ \citenamefont
  {Kobayashi}}]{PhysRevB.105.205123}%
  \BibitemOpen
  \bibfield  {author} {\bibinfo {author} {\bibfnamefont {D.}~\bibnamefont
  {Ohki}}, \bibinfo {author} {\bibfnamefont {K.}~\bibnamefont {Yoshimi}},\ and\
  \bibinfo {author} {\bibfnamefont {A.}~\bibnamefont {Kobayashi}},\ }\bibfield
  {title} {\bibinfo {title} {{Interaction-induced quantum spin Hall insulator
  in the organic Dirac electron system
  $\ensuremath{\alpha}\text{\ensuremath{-}}{\text{(BEDT-TSeF)}}_{2}{\mathrm{I}}_{3}$}},\
  }\href {https://doi.org/10.1103/PhysRevB.105.205123} {\bibfield  {journal}
  {\bibinfo  {journal} {Phys. Rev. B}\ }\textbf {\bibinfo {volume} {105}},\
  \bibinfo {pages} {205123} (\bibinfo {year} {2022})}\BibitemShut {NoStop}%
\bibitem [{\citenamefont {Ohki}\ \emph {et~al.}(2020)\citenamefont {Ohki},
  \citenamefont {Yoshimi},\ and\ \citenamefont
  {Kobayashi}}]{PhysRevB.102.235116}%
  \BibitemOpen
  \bibfield  {author} {\bibinfo {author} {\bibfnamefont {D.}~\bibnamefont
  {Ohki}}, \bibinfo {author} {\bibfnamefont {K.}~\bibnamefont {Yoshimi}},\ and\
  \bibinfo {author} {\bibfnamefont {A.}~\bibnamefont {Kobayashi}},\ }\bibfield
  {title} {\bibinfo {title} {{Transport properties of the organic Dirac
  electron system
  $\ensuremath{\alpha}\text{\ensuremath{-}}{(\mathrm{BEDT}\text{\ensuremath{-}}\mathrm{TSeF})}_{2}{\mathrm{I}}_{3}$}},\
  }\href {https://doi.org/10.1103/PhysRevB.102.235116} {\bibfield  {journal}
  {\bibinfo  {journal} {Phys. Rev. B}\ }\textbf {\bibinfo {volume} {102}},\
  \bibinfo {pages} {235116} (\bibinfo {year} {2020})}\BibitemShut {NoStop}%
\bibitem [{\citenamefont {Yoshimi}\ \emph {et~al.}(2023)\citenamefont
  {Yoshimi}, \citenamefont {Misawa}, \citenamefont {Tsumuraya},\ and\
  \citenamefont {Seo}}]{yoshimi2022}%
  \BibitemOpen
  \bibfield  {author} {\bibinfo {author} {\bibfnamefont {K.}~\bibnamefont
  {Yoshimi}}, \bibinfo {author} {\bibfnamefont {T.}~\bibnamefont {Misawa}},
  \bibinfo {author} {\bibfnamefont {T.}~\bibnamefont {Tsumuraya}},\ and\
  \bibinfo {author} {\bibfnamefont {H.}~\bibnamefont {Seo}},\ }\bibfield
  {title} {\bibinfo {title} {Comprehensive ab initio investigation of the phase
  diagram of quasi-one-dimensional molecular solids},\ }\href
  {https://doi.org/10.1103/PhysRevLett.131.036401} {\bibfield  {journal}
  {\bibinfo  {journal} {Phys. Rev. Lett.}\ }\textbf {\bibinfo {volume} {131}},\
  \bibinfo {pages} {036401} (\bibinfo {year} {2023})}\BibitemShut {NoStop}%
\bibitem [{\citenamefont {Varga}\ \emph {et~al.}(2003)\citenamefont {Varga},
  \citenamefont {Wilkinson},\ and\ \citenamefont {Angel}}]{FC70_FC77}%
  \BibitemOpen
  \bibfield  {author} {\bibinfo {author} {\bibfnamefont {T.}~\bibnamefont
  {Varga}}, \bibinfo {author} {\bibfnamefont {A.~P.}\ \bibnamefont
  {Wilkinson}},\ and\ \bibinfo {author} {\bibfnamefont {R.~J.}\ \bibnamefont
  {Angel}},\ }\bibfield  {title} {\bibinfo {title} {{Fluorinert as a
  pressure-transmitting medium for high-pressure diffraction studies}},\ }\href
  {https://doi.org/https://doi.org/10.1063/1.1611993} {\bibfield  {journal}
  {\bibinfo  {journal} {Rev. Sci.Instrum.}\ }\textbf {\bibinfo {volume} {74}},\
  \bibinfo {pages} {4564} (\bibinfo {year} {2003})}\BibitemShut {NoStop}%
\end{thebibliography}%
 %TC:endignore
\end{document}